\documentclass[11pt]{article}

\usepackage[]{acl}

\usepackage{times}
\usepackage{latexsym}
\usepackage{amsmath}
\usepackage{amssymb}
\usepackage[T1]{fontenc}
\usepackage[utf8]{inputenc}
\usepackage{CJKutf8}

\usepackage{microtype}
\usepackage{inconsolata}
\usepackage{graphicx}
\usepackage{seqsplit}
\usepackage{booktabs}
\usepackage{multirow}
\usepackage{float}
\usepackage{enumitem}
\usepackage{xcolor}
\usepackage{pifont}

\usepackage{calc}
\usepackage{placeins}
\usepackage{stfloats}
\usepackage{multicol}

\usepackage{longtable}
\usepackage{listings}
\usepackage{tipa}

\usepackage{xcolor}
\usepackage{colortbl}

\usepackage{tabularx}
\usepackage{multirow}

\usepackage{multicol}
\usepackage[most]{tcolorbox}
\tcbuselibrary{breakable}  
\tcbset{
  promptbox/.style={
    colback=gray!10,
    colframe=gray!40,
    boxrule=0.5pt,
    arc=3pt,
    left=6pt, right=6pt, top=4pt, bottom=4pt,
    fontupper=\small\ttfamily,
    width=\columnwidth,
    valign=top,
    halign=left,
  }
}

\usepackage{fancyvrb}

\lstdefinestyle{jsonblock}{
    basicstyle=\scriptsize\ttfamily,
    backgroundcolor=\color{gray!10},
    frame=single,
    framerule=0.4pt,
    rulecolor=\color{gray!40},
    columns=flexible,
    breaklines=true,
    showstringspaces=false,
    keepspaces=true
}

\lstdefinestyle{artifactid}{
    basicstyle=\scriptsize\ttfamily,
    backgroundcolor=\color{gray!10},
    frame=single,
    framerule=0.4pt,
    rulecolor=\color{gray!40},
    columns=flexible,
    breaklines=true,
    showstringspaces=false,
    keepspaces=true
}

\newcommand{\sym}[1]{#1}

\usepackage{titlesec}
\titlespacing*{\section}{0pt}{6pt}{4pt}
\titlespacing*{\subsection}{0pt}{4pt}{2pt}
\titlespacing*{\subsubsection}{0pt}{4pt}{2pt}
\titlespacing*{\paragraph}{0pt}{4pt}{1em}

\usepackage[disable]{todonotes}
\newcommand{\shama}[2][]
{\todo[color=cyan!15, inline, #1]{\textbf{SG:} #2}}

\newcommand{\hoang}[2][]
{\todo[color=yellow!15, inline, #1]{\textbf{HN:} #2}}

\title{CoSE-E: A Benchmark for Code-switched Speech Evaluation in Enterprise Settings}

\author{
 \textbf{Shama Gupta}$^{1}$,
 \textbf{Hoang Nguyen}$^{1}$,
 \textbf{Chelsea Huang}$^{2}$\thanks{Work done during internship at ServiceNow}
\\
\textbf{Lindsay Brin}$^{1}$,
 \textbf{Fanny Riols}$^{1}$,
\\
$^1$ServiceNow AI Research\quad $^2$ Qualcomm Technologies, Inc.\quad
\\
 \small{
   \textbf{Correspondence:} \href{mailto:shama.gupta@servicenow.com}{shama.gupta@servicenow.com}
 }
}

\def \benchmark{CoSE-E}
\def \csspanish{ES/EN}
\def \csgerman{DE/EN}
\def \csfrench{FR/EN}
\def \csfrenchcanadian{FR-CA/EN}
\def \cschinese{ZH/EN}

\def \assembly{AssemblyAI}
\def \deepgram{Nova-3}
\def \elevenlabs{Scribe-v2}
\def \gemini{Gemini-3-Flash}
\def \voxtralsmall{Voxtral}
\def \parakeet{Parakeet}
\def \qwen3{Qwen3-Omni}
\def \whisper{Whisper}
\def \gemma{Gemma4-31B IT}
\def \gpt{GPT-4.1}
\def \gptfive{GPT-5}

\begin{document}
\maketitle

\begin{abstract}
Code-switching (CS), a seamless alternation between languages within a single utterance, remains a critical challenge in automatic speech recognition (ASR). While prior works focus on conversational CS-ASR, enterprise settings demand evaluation of operational impact beyond edit-distance errors: how code-switching transcription errors propagate to downstream voice agent task failures. In this work, we propose (1) a CS-ASR synthetic benchmark and multidimensional evaluation framework tailored to enterprise domains, (2) systematic evaluation of frontier ASR systems across 5 language pairs, (3) diagnostic analysis of the additional transcription errors that code-switching introduces across language pairs and models. We release \textsc{CoSE-E} to support enterprise-focused CS-ASR evaluation for multilingual voice agents in enterprise deployment \footnote{\url{https://huggingface.co/datasets/ServiceNow-AI/asr_codeswitched}}. 
\end{abstract}

\section{Introduction}

Over half of the world's population is bilingual or multilingual~\cite{grosjean1982life}, and for many multilingual speakers, code-switching — seamlessly alternating between languages — is a natural communicative strategy across social and professional contexts~\cite{myersscotton1993duelling}. Still, CS-ASR remains a major challenge for current models~\cite{agro2025code,liu2026code}, and existing benchmarks~\cite{lovenia2022ascend,xie2026switchlingua} focus on conversational settings where code-switching is driven by social identity and stylistic preference~\cite{poplack1980typology,bullock2009cambridge,dogruoz2021survey}. However, enterprise environments — customer service, IT support, business process automation — exhibit fundamentally different patterns: speakers code-switch to navigate domain-specific terminology, clarify technical concepts, and accommodate mixed-language knowledge bases. Enterprise-specific CS corpora remain scarce, leaving ASR systems inadequately evaluated in the operational contexts where transcription failures are most consequential.

Evaluation methodology compounds the problem. Standard edit-distance metrics such as WER, CER, and MER~\cite{li2012code} lack cross-language standardization and, more importantly, measure transcription quality in isolation without quantifying the downstream impact of errors on tasks like ticket routing or policy retrieval~\cite{weng2020joint,shapira-etal-2025-measuring,bogavelli2026eva}. In enterprise voice-agent pipelines, where a misrouted ticket or misunderstood policy question has real operational consequences, transcription fidelity is the critical first link. Error propagation is not unique to code-switching, but we expect code-switching to amplify it: switch points are likely where transcription is least reliable, and the switched-in term is often the one that determines downstream action. However, this remains untested for enterprise code-switching.

We address both gaps by constructing \benchmark, a synthetic CS-ASR enterprise benchmark of 1,212 utterances spanning five language pairs — Spanish–English~(\csspanish), French–English~(\csfrench), Canadian French–English~(\csfrenchcanadian), German–English~(\csgerman), and Mandarin–English~(\cschinese) — grounded in HR and IT Service Management workflows. We evaluate eight ASR models using WER, Semantic Word Error Rate (SWER), and Answer Error Rate (AER), capturing both exact transcription accuracy and meaning preservation for downstream tasks. Our contributions are:

\begin{itemize}[nolistsep, noitemsep, topsep=0pt, partopsep=0pt, leftmargin=*]
\item \textbf{\benchmark}: A linguistically validated synthetic CS-ASR benchmark spanning five language pairs and authentic enterprise workflows.
\item \textbf{Comprehensive CS-ASR Evaluation}: Systematic evaluation of frontier ASR models across all five language pairs, quantifying code-switching robustness across surface-level and semantic metrics.
\item \textbf{Diagnostic Analysis}: Diagnostic analysis of the added transcription cost that code-switching imposes beyond monolingual baseline errors, revealing language-pair- and model-specific failure patterns.
\end{itemize}
\section{Related Work}

\paragraph{Code-Switched Speech Corpora}
Code-switched ASR has been studied across many language pairs. Wide-coverage benchmarks such as SwitchLingua~\cite{xie2026switchlingua} and CS-FLEURS~\cite{yan2025csfleurs} span dozens of pairs, while most modelling work is trained on smaller pair-specific corpora of conversational speech such as SEAME~\cite{lyu2010seame} and ASCEND~\cite{lovenia2022ascend} for Mandarin--English, with analogous resources for other pairs. These all report per-language WER/CER on the transcript, typically partitioned by \emph{matrix language}, the language setting the grammatical frame of an utterance, and \emph{embedded language}, the one filling content slots within that frame~\cite{myersscotton1993duelling}. Evaluation in these corpora is transcript-only, and they are drawn from social or conversational speech rather than enterprise interaction. \benchmark\ is, to our knowledge, the first benchmark designed around this setting.

\paragraph{Enterprise Code-Switching and Downstream Evaluation}
The corpora above evaluate transcripts in isolation, but ASR errors propagate to downstream tasks: \citet{lee2018spokensquad} show this for spoken QA, and \citet{oh2025confusion} document a related matrix-language bias in LLMs on code-switched input. Enterprise code-switching sharpens the problem: speakers stay in their matrix language and switch into English for domain-specific terms---\textit{VPN}, \textit{payroll cycle}, \textit{ticket}---that have no settled equivalent~\cite{broersma2006triggered,myslin2015language}, so errors concentrate on a small set of high-information words rather than scattering broadly. The closest prior benchmark~\cite{abdoli2026commercial} covers Arabic--English, Persian--English, and German--English on commercial ASR, but lacks Mandarin--English as a distinct-script pair, native-speaker review, and end-to-end downstream evaluation.

\begin{figure*}[t]
  \centering
  \includegraphics[width=0.8\textwidth, trim={0cm 0cm 0cm 0cm}, clip]{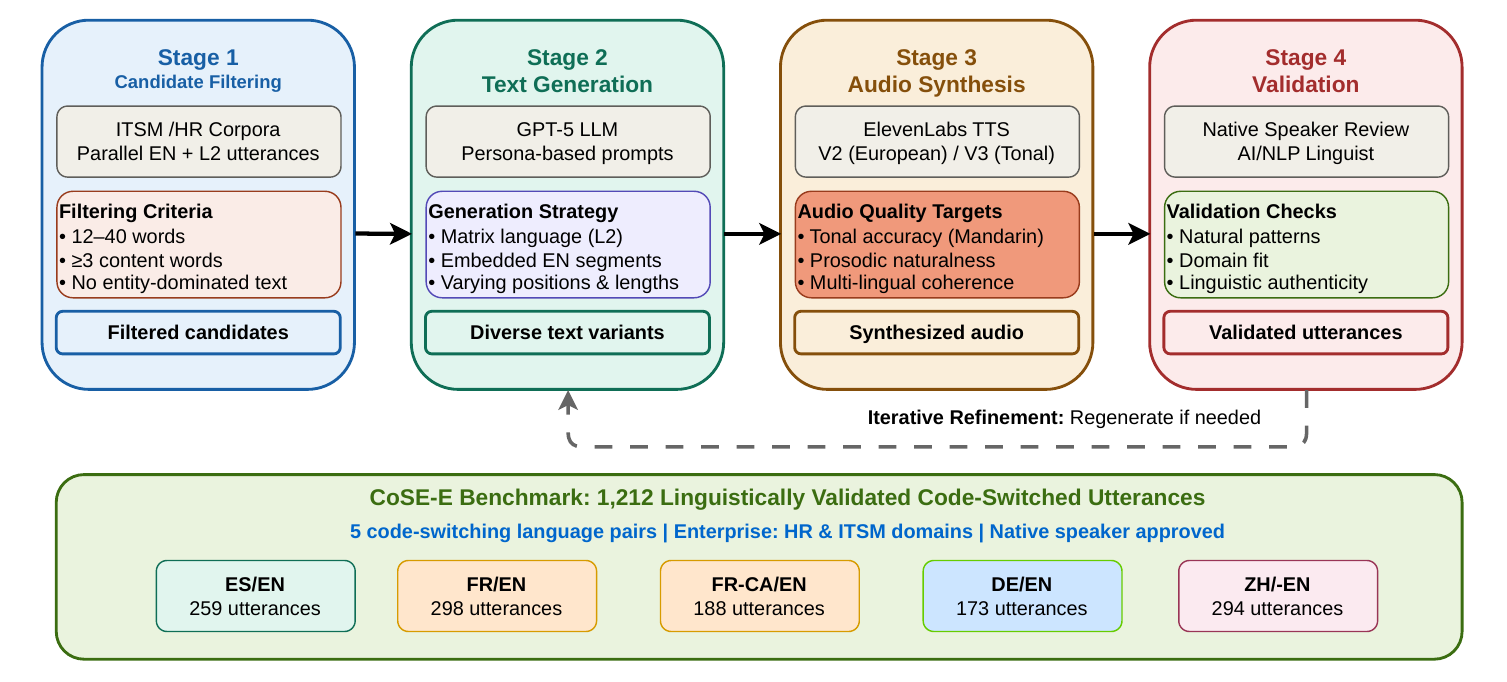}
  \caption{\textbf{Overview of \benchmark~Generation Pipeline}}
  \label{fig:generation_overview}
\end{figure*}

\section{Benchmark Design}
\begin{table*}[t]
\centering
\resizebox{\textwidth}{!}{%
\begin{tabular}{l l c c r r r r r r r}
\toprule
Language Pair & Dataset & Matrix Language & Embedded Language & \# utts & words/utt & chars/utt & switches/utt & CMI & switch density & \% mono ($\downarrow$) \\
\midrule
\multirow{3}{*}{\textbf{\csspanish}} & \textbf{\benchmark} & ES (64.2\%) & EN (35.8\%) & 259 & 24.60 & 114.76 & 6.91 & 0.32 & 0.29 & 0.00 \\
& CS-FLEURS  & ES (80.7\%) & EN (19.2\%) & 1576 & 26.01 & 129.33 & 5.90 & 0.16 & 0.24 & 3.24 \\
& SWITCHLINGUA & ES (64.5\%) & EN (35.5\%) & 8110 & 54.95 & 281.99 & 6.71 & 0.32 & 0.13 & 0.95 \\
\midrule
\multirow{4}{*}{\textbf{\csfrench}} & \textbf{\benchmark} & FR (57.2\%) & EN (42.8\%) & 298 & 21.56 & 98.54 & 4.95 & 0.33 & 0.24 & 0.00 \\
& \textbf{\benchmark~(CA)} & FR-CA (73.5\%) & EN (26.5\%) & 188 & 20.82 & 98.53 & 5.74 & 0.26 & 0.29 & 0.00 \\
& CS-FLEURS & FR (72.2\%) & EN (27.7\%) & 1315 & 25.46 & 129.82 & 6.17 & 0.19 & 0.25 & 3.42 \\
& SWITCHLINGUA & FR (64.4\%) & EN (35.6\%) & 8070 & 52.41 & 273.66 & 5.66 & 0.31 & 0.11 & 1.98 \\
\midrule
\multirow{3}{*}{\textbf{\csgerman}} & \textbf{\benchmark} & DE (63.7\%) & EN (36.3\%) & 173 & 21.12 & 109.32 & 4.99 & 0.30 & 0.25 & 0.00 \\
& CS-FLEURS  & DE (77.9\%) & EN (21.8\%) & 1507 & 22.20 & 130.02 & 5.60 & 0.20 & 0.27 & 3.78 \\
& SWITCHLINGUA & DE (62.0\%) & EN (22.1\%) & 15330 & 47.92 & 270.17 & 5.92 & 0.33 & 0.13 & 0.65 \\
\midrule
\multirow{3}{*}{\textbf{\cschinese}} & \textbf{\benchmark} & ZH (89.6\%) & EN (10.4\%) & 294 & 21.20 & 44.67 & 2.86 & 0.10 & 0.15 & 0.00 \\
& SEAME & EN (71.5\%) & ZH (28.3\%) & 3003 & 9.18 & 30.38 & 1.20 & 0.12 & 0.14 & 57.71 \\
& ASCEND & ZH (80.2\%) & EN (19.1\%) & 373 & 2.19 & 25.65 & 0.40 & 0.12 & 0.21 & 67.29 \\
\bottomrule
\end{tabular}%
}
\caption{\textbf{Linguistic characteristics of contemporary code-switching benchmarks}. \textbf{\benchmark}~demonstrates consistent code-switching patterns across languages (CMI 0.26–0.33, switch density 0.13–0.29) with no monolingual utterances, unlike baseline corpora which contain significant monolingual content (3.2–67.3\%). Column abbreviations: \# utts = \textit{utterance count}; words/utt = \textit{average words per utterance}; chars/utt = \textit{average characters per utterance}; switches/utt = \textit{code-switches per utterance}; CMI = \textit{code-mixing index}; switch density = \textit{proportion of switching frames}; \% mono = \textit{percentage of monolingual content}. All values reported to two decimal places.}
\label{tab:dataset_stats}
\end{table*}

\subsection{Data Pipeline}
Whether terms like \textit{laptop}, \textit{payroll}, and \textit{ticket} constitute 
code-switches or established loanwords is often ambiguous, as the boundary is 
speaker- and register-dependent~\cite{broersma2006triggered,myslin2015language}. 
For an ASR system and its downstream voice agent, however, this distinction is 
immaterial---both present the same recognition challenge: English-origin content 
words embedded in a non-English frame that must be transcribed correctly for the 
request to succeed. To evaluate these unified challenges, we introduce 
\textbf{\textsc{\benchmark}}, a synthetic code-switched speech benchmark grounded 
in two enterprise domains (HR and ITSM) covering five language pairs: 
\csspanish,~\csfrench,~\csfrenchcanadian,~\csgerman,~\cschinese. The pipeline 
proceeds in four stages, as shown in Figure \ref{fig:generation_overview}.

\paragraph{Stage 1: Parallel Data Selection \& Filtering.} We begin with an 
internal corpus of parallel ITSM and HR utterances spanning across English and each 
of the five non-English languages. Our filtering strategy enforces three major constraints: (1) \textbf{Length}: we retain only utterances between 12 and 40 words as they are long enough to contain natural code-switching, but short enough to represent realistic conversational turns, (2) \textbf{Entity Density}: we exclude utterances whose content is dominated by non-lexical entities (emails, phone numbers, IDs, URLs), where language alternation is structurally forced rather than pragmatically motivated. Entities themselves are retained throughout the benchmark and are a central object of our analysis (see \S\ref{sec:critical_entities}); the filter removes only utterances in which we believe entity strings crowd out the lexical material needed to study switching. (3) \textbf{Generation Feasibility}: we require at least three switchable content words (nouns, verbs, adjectives) per utterance. This is a constraint on our synthesis pipeline: utterances with sparse lexical material admit few candidate switch sites, causing the generation model to produce near-identical or degenerate variants. The threshold ensures each source utterance supports a range of switch points rather than a single forced one.

\paragraph{Stage 2: Code-Switching Generation.} 
We generate diverse code-switched text variants by leveraging LLMs with unconstrained persona prompts following insight from prior works that unconstrained persona-based prompting produces more natural switches than constrained strategies \cite{yan2025csfleurs,xie2026switchlingua}. The complete generation prompts are provided in Appendix \ref{app:prompts-datagen}.

\paragraph{Stage 3: Audio Synthesis.} We convert generated text utterances to spoken form through two sequential steps. First, a normalization and verbalization pass to convert text to phonetic representation, ensuring consistent pronunciation across variants. This pass standardizes how numbers, symbols, and special characters are rendered, so that the synthesis is accurate. The normalization rules and LLM verbalization 
prompts are detailed in Appendix~\ref{app:normalization}. Second, we synthesize audio using ElevenLabs Multilingual V2 for the four European language pairs (DE, ES, FR, FR-CA) and ElevenLabs V3 for ZH/EN, which provides superior tonal accuracy and prosodic naturalness critical for ZH. We select one male and one female voice for each language, both verified by native speakers, resulting in 10 voices across the benchmark with balanced gender representation (50\% male, 50\% female utterances). This two-stage approach ensures both linguistic fidelity and acoustic authenticity.

\paragraph{Stage 4: Data Quality Validation.} Quality is essential for a benchmark intended to rigorously evaluate the systems' robustness in real-world code-switching enterprise contexts. Thus, we implement a systematic validation by professionally trained linguists with two combined critical competencies: native fluency in each matrix language and familiarity with code-switching patterns and linguistic norms specific to enterprise contexts. Utterances flagged for unnatural switching patterns or phonetic mispronunciations are either excluded or returned to Stage 2 for regeneration and re-review. 

\paragraph{Dataset Overview} The resulting dataset contains 1,212 records 
distributed across five language pairs: \csspanish~(259), \csfrench~(298), 
\csfrenchcanadian~(188), \csgerman~(173), and \cschinese~(294), as summarised 
in Table~\ref{tab:dataset_stats}. A critical distinction of \benchmark~is its 
enforced zero-monolingual constraint (\textit{\%mono=0.00\%}) across all language 
pairs, guaranteeing every record is a genuine code-switched utterance. This 
contrasts with existing benchmarks, which contain substantial monolingual content 
--- particularly Mandarin--English corpora such as SEAME (57.7\%) and ASCEND 
(67.3\%). \benchmark~also maintains consistent code-switching depth across 
European language pairs, as measured by three jointly-optimised metrics: switches 
per utterance (4.9--6.9), code-mixing index (0.26--0.33), and switch density 
(0.13--0.29). The Mandarin--English pair exhibits lower values (switches/utt: 
2.86, CMI: 0.10, switch density: 0.15), reflecting the predominantly 
\emph{insertional} nature of mainland-China Mandarin--English code-switching, 
where speakers embed single English content words or short phrases into a Mandarin 
frame~\cite{zhong2024mandarin}. This pattern is especially prevalent in enterprise 
settings, producing fewer, shorter switches per utterance than the European pairs. 

\subsection{Evaluation Metrics}
\label{sec:eval_metrics}
WER assumes a single canonical reference transcript, an assumption code-switching undermines. Transliteration, mixed script, borrowing-versus-switching ambiguity, and casing all admit multiple valid surface forms for the same utterance, causing normalization and segmentation choices to exert outsized influence on results. In enterprise CS-ASR the measurement requirements sharpen further: success of a voice agent hinges not on surface fidelity but on correctly transcribing critical entities and technical terms while preserving semantic meaning—failures that WER alone cannot capture. Accordingly, we propose Answer Error Rate (AER) as the standard evaluation metric for code-switched ASR in enterprise domains, and adopt it as the primary metric throughout this paper. We report AER alongside WER and SWER for comparability with prior works.

\paragraph{Answer Error Rate (AER).} Introduced by \citet{pulikodan2025approach}, AER measures task-level fidelity: it captures whether transcription errors impair a downstream LLM's ability to understand the important details. A three-role LLM pipeline (Figure~\ref{fig:aer_pipeline} in Appendix \ref{app:aer_stochasticity}) operationalizes this. A \emph{question generator} produces k=3 comprehension questions from the reference transcript, targeting the critical entities that drive voice-agent task success---identifiers, technical terms, and action-bearing tokens \cite{bogavelli2026eva}. We have three human annotators confirm that questions  have a grounded answer in the reference utterance (Appendix \ref{app:human_validation}). An \emph{answerer} responds to each question twice—once from the reference transcript and once from the hypothesis. An \emph{alignment judge} then compares the two answer sets question by question; AER is the fraction of questions whose hypothesis-derived answer diverges from the reference-derived one. We test both components for stochasticity in Section~\ref{app:aer_stochasticity}.

Two properties make AER the right primary metric for code-switched evaluation.

(1) \textit{Normalization-invariant.} Scoring operates over answers rather than strings, so normalization choices that destabilize WER under code-switching never enter the computation.

(2) \textit{Consequence-aware scoring.} AER's questions target contentful spans—entities, technical terms, alphanumerics—the highest-information tokens in enterprise speech and the ones code-switching most often corrupts (\S\ref{sec:critical_entities}). By design, information-dense tokens are scored directly rather than averaged away as they would be under WER's uniform weighting. We confirm this directly in \S\ref{sec:critical_entities}: AER's answer changes precisely when code-switching destroys such a token, and remains stable when the token is only reformatted. AER is thus a task-aligned proxy rather than a substitute for measured task completion, which we leave to future work.

We instantiate the question generator and alignment judge with Gemma-4-31B and the answerer with GPT-4.1 (full prompts in Appendix~\ref{app:prompts}). 

\paragraph{Complementary metrics.} We report two additional metrics to enable comparison with prior CS- and multilingual-ASR work. \textbf{Word Error Rate (WER)} is the surface baseline. \textbf{Semantic WER (SWER)} captures semantically meaningful errors, crediting meaning-preserving surface differences and penalizing meaning-altering ones; 
\hoang{The rest of this statement is the implementation details and should belong to the experimental setup instead (Section 4). The following statement repeats the statement in the prior sentence.}
we adapt Pipecat's open-source STT benchmark\footnote{\url{https://github.com/pipecat-ai/stt-benchmark}}. For ZH/EN we additionally report Character Error Rate (CER), the standard evaluation unit for Chinese ASR, along with WER.
\begin{table*}[t]
\centering
\scriptsize
\setlength{\tabcolsep}{3.5pt}
\begin{tabular}{l ccc ccc ccc ccc c|ccc}
\toprule
& \multicolumn{3}{c}{\csspanish} & \multicolumn{3}{c}{\csfrench} & \multicolumn{3}{c}{\csfrenchcanadian} & \multicolumn{3}{c}{\csgerman} & \multicolumn{4}{c}{\cschinese} \\
\cmidrule(lr){2-4}\cmidrule(lr){5-7}\cmidrule(lr){8-10}\cmidrule(lr){11-13}\cmidrule(lr){14-17}
Model & WER & SWER & AER & WER & SWER & AER & WER & SWER & AER & WER & SWER & AER & WER & CER & SWER & AER \\
\midrule
Scribe-V2        & \textbf{.022} & \textbf{.004} & .033 & \textbf{.031} & \textbf{.006} & \textbf{.051} & \textbf{.041} & \textbf{.005} & \textbf{.030} & \textbf{.027} & \textbf{.002} & \textbf{.021} & .073 & \textbf{.031} & \textbf{.006} & .042 \\
Gemini-3-Flash   & .028 & .005 & .031 & .040 & .009 & .054 & .055 & .008 & .043 & .046 & .003 & .023 & .090 & .041 & .009 & .059 \\
AssemblyAI       & .029 & \textbf{.004} & .033 & .039 & .009 & .062 & .052 & .006 & .034 & .048 & .003 & .023 & .093 & .046 & .010 & .057 \\
Qwen3-Omni       & .042 & \textbf{.004} & \textbf{.027} & .061 & .010 & .055 & .071 & .009 & .062 & .063 & .004 & .033 & \textbf{.040} & .039 & \textbf{.006} & \textbf{.041} \\
Voxtral          & .049 & .005 & .036 & .060 & .012 & .068 & .059 & .014 & .074 & .081 & .003 & .027 & --   & --   & --   & --   \\
Parakeet         & .117 & .027 & .075 & .075 & .024 & .084 & .099 & .025 & .069 & .054 & .005 & .039 & --   & --   & --   & --   \\
Nova-3           & .042 & .013 & .088 & .052 & .019 & .065 & .072 & .019 & .080 & .069 & .012 & .071 & --   & --   & --   & --   \\
Whisper          & .161 & .030 & .073 & .434 & .061 & .111 & .535 & .029 & .089 & .615 & .045 & .092 & 1.494 & 2.516 & .514 & .515 \\
\bottomrule
\end{tabular}
\caption{\textbf{WER, SWER, and AER by model and language pair.} \cschinese~WER is jieba word-level (mean of per-utterance WER, as for all pairs) and broadly comparable to the Latin-script pairs; \cschinese~CER is character-level and not magnitude-comparable. All \cschinese~outputs were normalized Traditional$\rightarrow$Simplified (OpenCC) before WER/CER/SWER scoring. Best per column in bold; ties at displayed precision (3 decimals) are both bolded.}
\label{tab:main}
\end{table*}
\section{Results}
\label{sec:results_intro}

We evaluate eight frontier models covering both proprietary  and open-weight across five language pairs: AssemblyAI Universal-3.5-Pro~\footnote{\url{https://www.assemblyai.com/universal-3-pro}}, Deepgram Nova-3 Multilingual~\footnote{\url{https://developers.deepgram.com/docs/live-streaming-audio}}, ElevenLabs Scribe-v2, Google Gemini-3 Flash~\footnote{\url{https://ai.google.dev/gemini-api/docs/models/gemini-3-flash-preview}}, Voxtral-Small-24B~\cite{liu2025voxtral}, Parakeet-TDT-0.6b-v3~\cite{sekoyan2025canary}, Qwen3-Omni-Instruct~\cite{xu2025qwen3}, and Whisper-Large-v3-Turbo~\cite{radford2023robust}. To simulate the realistic enterprise deployment settings, all models use native auto-language detection without explicit language parameter \footnote{Language parameter refers to language-specific names, codes, abbreviations, depending on the support of each provider}, as detailed in Appendix \ref{app:evaluated_models_settings}. 

For SWER judgment and AER evaluation, we instantiate the question generator and alignment judge with Gemma-4-31B-IT~\cite{team2026gemma}(max tokens=12000, temperature=1.0, top p=0.95, top k=64) and the answerer with GPT-4.1~\footnote{\url{https://developers.openai.com/api/docs/models/gpt-4.1}} (temperature=0). The hyperparameters for all evaluated ASR systems and judge models are 
detailed in Appendix~\ref{app:evaluated_models_settings}.
\subsection{Model performance across metrics}
\label{sec:model_performance}
Model performance separates into tiers (Table \ref{tab:main}), but a model's tier sometimes changes by metric. We use a Friedman test to confirm all eight models differ significantly on every metric (WER: $\chi^2(7, N{=}918){=}1464.1$; SWER: $612.2$; AER: $200.1$; all $p<.001$; table \ref{tab:stats} in Appendix \ref{app:model_stats}). On WER, the field splits three ways: Scribe-v2, Gemini-3-Flash, and AssemblyAI cluster at the top (WER $0.02$--$0.05$ across pairs); Qwen, Voxtral, Nova-3, and Parakeet occupy the middle; and Whisper ranks far below, translating rather than transcribing code-switched audio under automatic language detection (Appendix \ref{app:whisper_ablation}). But AER reshuffles the middle tier. Qwen ranks sixth on \csspanish~WER ($0.042$) yet first on AER ($0.027$)---ahead of every model, including Scribe. The gap is substantial: Qwen's AER is $18\%$ lower than Scribe's on that pair (Table \ref{tab:main}). A mid-pack model on WER can be top-ranked in AER.

Within the top tier, there is no clear winner. The top four---Scribe, Gemini, AssemblyAI, and Qwen---are all mutually non-significant on AER after Holm correction, and within-metric concordance is low ($W{=}0.23$ for WER, $0.03$ for AER; Table~\ref{tab:stats}), confirming that models trade places from clip to clip. The best system therefore depends on the target language pair: Scribe leads almost everywhere on WER, but Qwen takes the lowest AER on both \csspanish~and \cschinese, while Gemini takes second-lowest AER on \csspanish~despite ranking third on WER there. Notably, both are large audio language models rather than ASR systems; we hypothesize that stronger language modeling compensates for surface transcription errors by preserving the semantic content that AER captures.

AER separates models where surface metrics cannot, because it isolates semantic failures from normalization choices. All three metrics agree on the coarse three-tier ranking (Kendall's $W = 0.963$), but diverge within tiers: on European pairs, AER still distinguishes models while WER and SWER collapse (within-metric $W$ falls from $0.23$ to $0.03$) Table \ref{tab:stats}. Chinese reveals why. Gemini's WER improves 46\% after Traditional-to-Simplified conversion ($0.167 \to 0.090$), yet AER is unchanged ($0.059$)---the judge recovers answers regardless of script (shown in Table \ref{tab:t2s_ablation} in \ref{app:normalization_effect}). Whisper shows the same pattern: CER of $2.52$, but AER bounded at $0.52$ (Table \ref{tab:main}). Surface metrics conflate encoding choices with transcription errors; AER does not, making it the more reliable metric to predict task success for enterprise use cases.

\subsection{Model performance across language pairs}
\label{sec:rq2}

\begin{table}
\centering
\small
\begin{tabularx}{\columnwidth}{Xccc}
\toprule
\textbf{Language pair} & \textbf{WER} & \textbf{SWER} & \textbf{AER} \\
\midrule
EN/DE    & 2.62         & \textbf{1.38} & \textbf{1.50} \\
EN/ES    & \textbf{1.75}& 2.00          & 2.25 \\
EN/FR-CA & 3.62         & 3.13          & 2.88 \\
EN/FR    & 2.00         & 3.50          & 3.38 \\
\midrule
Friedman $\chi^2(3)$ & 10.05 & 13.95 & 9.45 \\
$p$                  & .018  & .003  & .024 \\
Kendall's $W$        & 0.42  & 0.58  & 0.39 \\
\bottomrule
\end{tabularx}
\caption{\textbf{Mean rank of each language pair per metric (4 pairs, 8 models).}
Rank~1 = easiest, rank~4 = hardest; averaged across all 8 models. Bold = easiest pair per metric. Friedman $\chi^2$ and Kendall's $W$ summarise whether pairs differ significantly and how consistently models agree on the ordering.}
\label{tab:rq2-mean-rank}
\end{table}
 
We ask how model performance differs by language pair, and whether
the ranking of pairs is the same across metrics. We test this on two
designs: a primary analysis of the four Latin-script pairs across all
eight models (shown in Table \ref{tab:rq2-mean-rank}, and a restricted analysis that adds \cschinese~on the five
models that support it (Scribe-V2, Gemini-3-Flash, AssemblyAI,
Whisper, Qwen-3-Omni), shown in Table \ref{tab:rq2-mean-rank-zh} in Appendix \ref{app:language_ranks_chinese}. On the primary design, a Friedman test finds
the clearest separation on SWER ($\chi^2(3){=}13.95$, $p{=}.003$;
$W{=}0.58$), a moderate one on WER ($\chi^2(3){=}10.05$, $p{=}.018$,
$W{=}0.42$), and a similar one on AER ($\chi^2(3){=}9.45$, $p{=}.024$,
$W{=}0.39$) (Table \ref{tab:rq2-mean-rank}). All three survive Holm correction, but the signal is
sharpest on SWER: semantic difficulty is the most consistently
detectable difference across language pairs.

On the meaning-aware metrics, \csgerman~is the easiest language pair and
\csfrench~the hardest. SWER and AER agree on the full ordering---\csgerman~
easiest, then \csspanish, \csfrenchcanadian, and \csfrench~(AER ranks $1.50$, $2.25$,
$2.88$, $3.38$; SWER nearly identical). WER produces a different
ordering: \csspanish~easiest ($1.75$), then \csfrench ($2.00$), \csgerman~
($2.62$), and \csfrenchcanadian~hardest ($3.62$) (Table \ref{tab:rq2-mean-rank}). The two orderings diverge specifically on German and French, which trade the easy and hard ends: EN/DE is second-hardest on WER but easiest on both meaning-aware metrics, while \csfrench~is second-easiest on WER but hardest on both. This reversal reflects the errors
each language produces: German accumulates surface word-form errors
(morphology, compounding, casing) that inflate WER but rarely touch
task entities, while French produces fewer errors that land on
content that matters. Because WER weighs the cheap German errors and consequential French errors alike, its ranking is misleading.

Chinese extends the same pattern to a larger scale. In the restricted
five-model design, \cschinese~is unambiguously the hardest pair on the
surface metrics---ranked hardest by every model on WER ($5.0$) and
nearly every model on SWER ($4.6$). On AER, however, it is no longer
hardest: \csfrench~edges past it ($4.4$ vs.\ $4.2$) (Table \ref{tab:rq2-mean-rank-zh}). Chinese produces many
surface errors, but they scatter across word forms, segmentation, and normalization 
rather than concentrating on consequential content. A pair with far higher surface error can thus be easier on the metric
that matters---the clearest evidence that surface scores alone
mischaracterize code-switch difficulty for enterprise evaluation.

\section{Discussion}
\subsection{What additional cost does codeswitching add compared to plain monolingual speech?}

\label{sec:rq3}
 
We define code-switching cost as the per-utterance error delta between the monolingual and code-switched conditions described in \S\ref{sec:results_intro}. Because a mixed-language utterance has no single correct language tag, we transcribe the code-switched condition language-agnostically and the monolingual condition with its unambiguous tag, matching how each would be configured in deployment.

For each combination of model, language pair, baseline (English or non-English monolingual), and metric (WER, SWER, AER), we pair utterances by record ID, keeping only records present in both runs. We manually reviewed each audio to ensure it matched the ground truth and discarded mismatched pairs. For each retained utterance $i$ we compute $\Delta_i = m(\text{cs}_i) - m(\text{mono}_i)$,
where $m$ is the metric; positive $\Delta$ indicates code-switching increased error, and negative $\Delta$ indicates code-switching decreased error. Since the deltas are zero-inflated and right-skewed, we test each mean against zero with a two-sided sign-flip permutation test (10{,}000 permutations). We control the family-wise error rate across
models within each language-pair $\times$ baseline $\times$ metric family
using Holm--Bonferroni, and report a 95\% bootstrap confidence interval on
each mean delta.

\subsubsection{Code-switching raises word-level error}
\label{sec:results-surface}
 
\begin{figure*}[t]
  \centering
  \includegraphics[width=\textwidth]{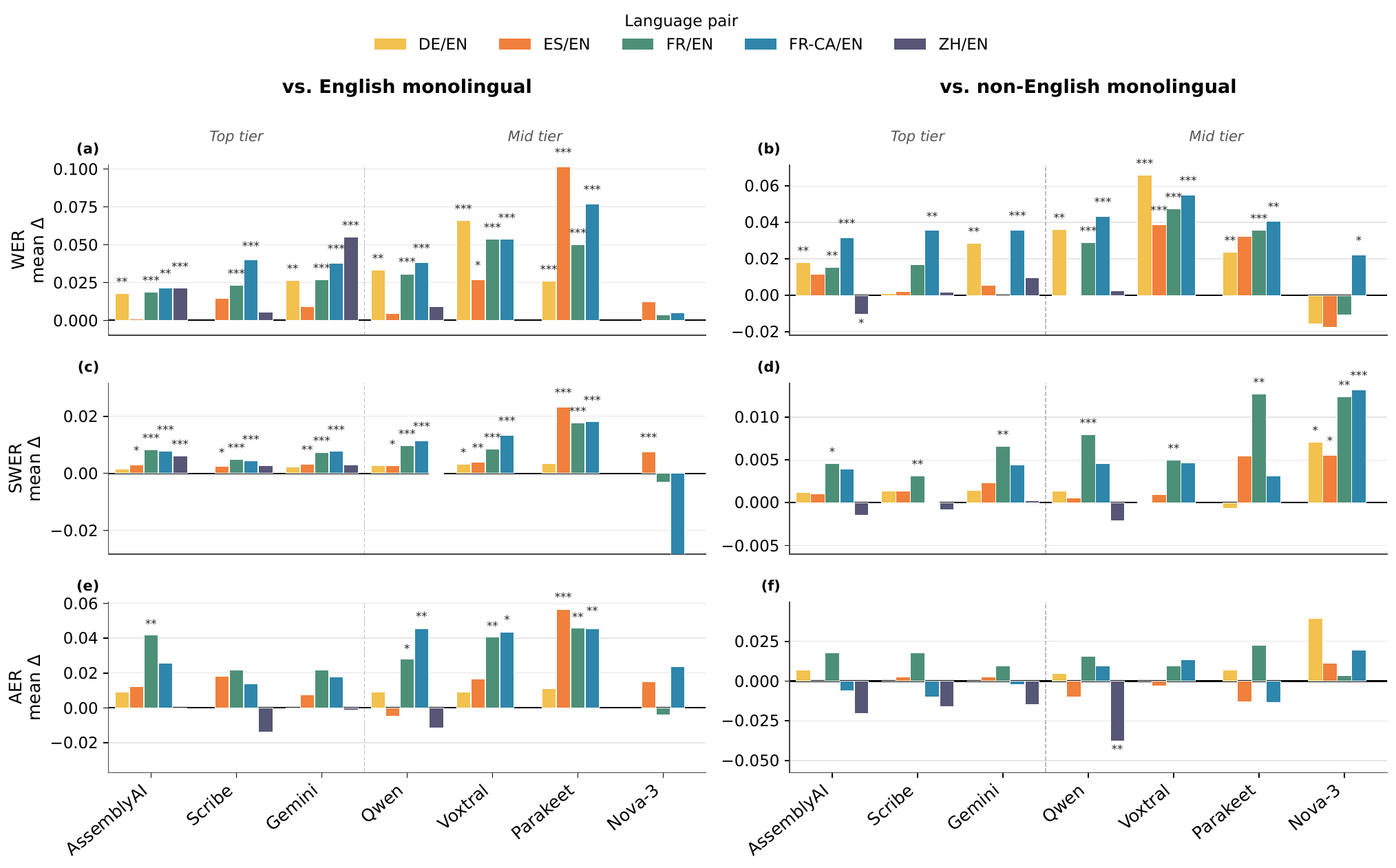}
  \caption{\textbf{Mean per-utterance code-switching cost},
    $\Delta = m(\text{cs}) - m(\text{mono})$; positive values indicate
    code-switching is harder. Rows: WER, SWER, AER. Columns: the two
    monolingual baselines. Bars are grouped by model (top/mid tier) and
    colored by language pair; whiskers are $95\%$ bootstrap CIs. Stars mark
    significance under a two-sided sign-flip permutation test, Holm-corrected
    within each metric\,$\times$\,pair\,$\times$\,baseline family
    ($^{*}p{<}.05$, $^{**}p{<}.01$, $^{***}p{<}.001$). Whisper is omitted; its
    deltas reach $+2.5$ (see Figure ~\ref{fig:cs_delta_whisper}).}
  \label{fig:cs_delta}
\end{figure*}

Code-switching increases word-level error for nearly every system and
language pair we evaluate. Under WER, the per-utterance delta is significantly positive in 39 of 62 model $\times$ language pair $\times$ baseline combinations ($63\%$; two-sided sign-flip permutation, Holm-corrected within each family), excluding Whisper, which fails in a categorically different way (see below). The effect is modest for the
strongest systems---AssemblyAI, Scribe, and Gemini stay within roughly
$+0.02$ to $+0.05$---and larger for the mid tier, where Voxtral and Parakeet
reach $+0.07$ to $+0.10$ on the harder pairs (Figure~\ref{fig:cs_delta}, top
row). This is consistent with prior work reporting that mixed-language speech
is difficult to transcribe: at the surface, code-switching is a real and
measurable cost.
 
Whisper fails completely rather than incrementally. Run without a specified language---as the bilingual condition requires---it does not transcribe the mixed utterance at all: it detects a single language and translates the rest of the speech into English. The output therefore diverges wholesale from the code-switched reference, with deltas reaching +2.5 WER (Figure \ref{fig:cs_delta_whisper}) on~\cschinese, an order of magnitude beyond any other system. This is a decoding-mode failure, not a larger version of the same code-switching cost, so we report Whisper separately (Appendix \ref{app:whisper_ablation}).
 
\subsubsection{Code-switching hurts some language pairs more than others}
\label{sec:results-gradient}
 
The code-switching cost varies across language pairs (Table~\ref{tab:cs_delta_full}). \csfrench~is the
most expensive and most consistent, significant across all three metrics
(WER 10/14, SWER 13/14, AER 4/14 model $\times$ baseline combinations), converging with our\shama{Will update these references to the Appendix if I add the Two part Model study there. if not, i will delete these references and sentences relating.} \S\ref{sec:two_part_model} finding that
switch count predicts error onset most strongly in \csfrench~
(\S\ref{sec:two_part_model}). \csgerman~is costly at the surface (WER 10/12) but
flat under SWER (2/12) and null under AER, indicating errors that are
surface-level and semantically recoverable. \csspanish~sits in between, which is consistent with our results from \S\ref{sec:rq2}.
 
\cschinese~ has little change between code-switched and monolingual baselines, which we hypothesize is due to a smaller switch count and CMI (Table \ref{tab:dataset_stats}).
 
\subsubsection{Code-switching leaves meaning largely intact---but
  the errors that survive are consequential}
\label{sec:results-collapse}

The measured cost of code-switching shrinks steadily as the metric moves
from verbatim transcription (WER) toward semantic fidelity (SWER) and
predicted task success (AER). The share of  model $\times$ language pair $\times$ baseline combinations with a significant
code-switching penalty falls from $63\%$ under WER to $50\%$ under SWER
and $15\%$ under AER (Table \ref{tab:cs_delta_full}), and the effect grows more concentrated at the
utterance level: the fraction of utterances left completely unchanged
rises from $35\%$ under WER to roughly $80\%$ under both SWER and AER. Most of the word errors that code-switching
introduces do not change the meaning of the utterance or the agent's
answer.

The $15\%$ that do reach AER, however, are not marginal cases---they are
utterances where the downstream answer changes, which in an enterprise
deployment means the voice agent would have been unable to complete its task successfully. By construction, a non-zero AER marks
a record where the hypothesis-derived answer diverges from the
reference-derived one (Appendix \ref{app:aer_error_examples}). Every model $\times$ language pair $\times$ baseline combination with a
significant AER penalty therefore represents a cluster of utterances where
code-switching would likely cause genuine task failure, not merely a noisier
transcript. The surface difficulty that prior work documents is real but
shallow; conversely, AER errors are the
opposite---rare, but consequential wherever it appears. 

\subsubsection{Code-switching errors that reach AER are critical-entity failures}
\label{sec:critical_entities}

What breaks when code-switching reaches the task level? In 57\% of
cases, the cause is a corrupted critical entity: an ID, hostname, URL,
or email that the agent needs intact to complete the request. On one
IT-support utterance, the hostnames \texttt{host-2854} and
\texttt{host-1375} are transcribed correctly in both the English and
French monolingual conditions but corrupted under code-switching by
every model (\texttt{host2win54}, \texttt{OS365}, \texttt{ostewin54}),
flipping AER from $0$ to $1$ each time (Appendix~\ref{app:aer_error_examples}).

This link is systematic. When code-switching introduces an entity
error, AER worsens $\sim$40\% of the time against a base rate of
$\sim$5\%---an eightfold increase that holds across every language pair
and both baselines (Table~\ref{tab:entity_propagation}). Conversely,
AER stays silent when an entity is merely reformatted: casing changes
(\texttt{sn\_safe\_story} $\rightarrow$ \texttt{SNSafeStory}) and
digits read aloud as words leave the answer unchanged because the
value is still recoverable.

\begin{table}[t]
\centering\small
\setlength{\tabcolsep}{5pt}
\begin{tabular}{lcccc}
\toprule
 & \multicolumn{2}{c}{P(AER worsens)} & & \\
\cmidrule(lr){2-3}
Baseline & entity err. & no err. & RR & OR \\
\midrule
English     & 0.44 & 0.054 & 8.3 & 14.0 \\
non-English & 0.39 & 0.052 & 7.6 & 11.8 \\
\bottomrule
\end{tabular}
\caption{\textbf{Entity-error propagation to task failure}. RR = risk ratio; OR = odds ratio. When code-switching
  introduces a critical-entity error, the chance that the downstream
  answer worsens rises roughly eightfold, near-identically against both
  monolingual baselines (per-pair RR $6$--$13$, all $p<10^{-30}$).
  Whisper excluded; records inner-joined by ID.}
\label{tab:entity_propagation}
\end{table}

How often code-switching mistranscribes an entity depends on the baseline. Against the
non-English baseline, entity errors are introduced less often
(9.9\% vs.\ 13.6\% of records) and aggregate entity error falls
rather than rises ($-0.02$ vs.\ $+0.05$), consistent with
\S\ref{sec:results-collapse}. But the propagation rate is unchanged (RR 
7.6 vs.\ 
8.3)---the link from entity corruption to simulated task failure holds regardless of baseline. WER counts reformatting and semantic destruction alike; AER isolates the errors that would actually cause an enterprise voice agent to fail.

\subsection{Do code-mixing index, switch count, and utterance length predict transcription errors?}
\label{sec:two_part_model}
We used a two-part model to test whether CMI, switch count, and utterance length predict transcription failure on codeswitched speech. Part A is a logistic regression on error occurrence (WER > 0); Part B is OLS on log(WER), fit only to erroneous utterances, predicting failure magnitude. We fit both per model and per language pair, giving up to seven independent replications of each effect (see Tables~\ref{tab:dataset-composition} and~\ref{tab:base-rate} for dataset and base-rate details). Our discussion centers on Part~A (Table~\ref{tab:part-a-or}); we report Part~B briefly and note that its length coefficient is partly definitional, as WER is itself normalized by utterance length.

\paragraph{CMI does not predict whether an error occurs.} Across nearly all model–language combinations, the odds ratio for CMI sits at roughly 1.0 and is non-significant, indicating that the overall balance of the two languages carries no signal for error onset once switches and length are controlled. Only 2 of 32 combinations reached significance with no consistent direction (Table~\ref{tab:part-a-or}). Thus, we find that the degree of mixing is not predictive of error occurrence.

\paragraph{Switch count predicts error onset, but mainly for \csfrench.} In \csfrench, 6 of 7 models showed significantly elevated odds of error with each additional switch (Nova-3 the only exception). This effect held while controlling for length (switch count and length were only moderately correlated, $r = 0.56$, VIF 1.46; see Tables~\ref{tab:dataset-composition} and~\ref{tab:collinearity}), so it reflects the switches themselves rather than longer utterances carrying more of them. Elsewhere the pattern was sparse: switch count was significant for only Whisper in \csgerman, only Gemini-3 in \csfrenchcanadian, and only Whisper in \csspanish, and for no model in \cschinese. Switch count is thus a robust driver of errors in one language pair and a scattered one in the rest.

\paragraph{Length confounds severity analysis.} In Part~B, $\log(n_{\text{words}})$ dominates error magnitude, but this partly reflects WER's construction, rather than code-switching effects. CMI and switch count show no consistent relationship to severity, with only isolated negative CMI effects for Whisper (\csgerman, \cschinese).

\section{Conclusion}

We introduced \textsc{\benchmark}, an enterprise-focused CS-ASR benchmark of 1,212 linguistically validated utterances spanning five language pairs, and evaluated eight frontier ASR systems across three metrics. Our evaluation revealed a three-tier performance structure, with Scribe-v2, Gemini-3-Flash, and AssemblyAI consistently leading.

Code-switching imposes a real but largely surface-level cost: 63\% of model $\times$ language pair $\times$ baseline combinations show a significant WER penalty, but this drops to $15\%$ under AER, where roughly $80\%$ of utterances show no change in downstream answer. The errors that do survive are disproportionately critical-entity failures---corrupted IDs, hostnames, and URLs---with an eightfold increase in task failure probability when such an entity is destroyed. AER proves more robust than WER for enterprise CS-ASR evaluation, resisting the normalization ambiguities that destabilize surface metrics across scripts and morphologies.

WER alone mischaracterizes CS-ASR readiness. A system with moderate WER may preserve all task-critical content, while one with lower WER may fail on exactly the entities that matter. We release \textsc{\benchmark} and support CS-ASR evaluation for multilingual voice agents in enterprise settings.
\section*{Limitations}
\label{app:limitations}

\paragraph{Synthetic audio.}
\textsc{CoSE-E} audio is synthesized with ElevenLabs Multilingual text-to-speech (V2 for the European pairs, V3 for Mandarin) rather than recorded from human speakers. It does not include telephony bandwidth, disfluencies, or background noise. Reported results are therefore an upper bound on deployment quality.

\paragraph{Single vendor bias} We deliberately chose ElevenLabs TTS engine for audio synthesis to ensure high-quality data for the benchmark. However, we acknowledge this introduces potential vendor bias since ElevenLabs Scribe V2 is among the evaluated ASR systems and may benefit from familiarity with ElevenLabs' TTS characteristics. Our future works seek to expand evaluation towards multiple TTS providers to warrant TTS-independent CS-ASR performance.

\paragraph{Coverage.}
The benchmark covers five language pairs, all with English as the embedded language, and two enterprise verticals (HR and ITSM). Other high-volume pairs (e.g., Hindi--English, Arabic--English, Tagalog--English), other verticals (e.g., finance, healthcare), are out of scope. Only Mandarin contributes a non-Latin script.

\paragraph{Corpus scope.}
All utterances are fully code-switched (\%mono $=0$) and restricted to intra-sentential insertional switching; alternational and inter-turn switching~\cite{muysken2000bilingual} are excluded. We also do not separate insertional switches from established English loanwords, labelling both as ``embedded English''; the two are consequently not analyzed separately in our error breakdowns. The Mandarin--English pair has lower switch density than the European pairs, consistent with the predominantly insertional character of mainland-China Mandarin--English code-switching~\cite{zhong2024mandarin}; we report Mandarin as a separate slice throughout, since cross-pair comparisons involving Mandarin conflate typology with model capability. We seek to extend our work towards more typologically diverse languages as explored in recent text-based multilingual research works \cite{nguyen2024cori,nguyen2025prompting,ploeger2026principled}.

\paragraph{AER harness.}
AER is an LLM-based question-answering proxy for voice-agent task success, not an end-to-end evaluation. Rankings may shift under a different judge configuration. Our stochasticity study (Appendix~\ref{app:aer_stochasticity}) bounds the noise floor at the reported scale but does not rule out systematic judge sensitivity. AER is also substantially more compute-intensive than WER and is intended for model selection rather than continuous monitoring.

\paragraph{Evaluation scope.}
Evaluation is zero-shot on off-the-shelf ASR and audio-LLM systems as of mid-2026. We do not measure the effect of targeted fine-tuning or contextual biasing on the CS-ASR distribution. Absolute numbers are a point-in-time snapshot; the qualitative pattern (non-zero code-switching cost and AER separation within surface-metric ties) is what we expect to generalize.

\bibliography{custom}
\UseRawInputEncoding
\newpage
\appendix
\section{Prompts}
\label{app:prompts}

This appendix lists the prompts used to build our dataset. Curly-brace tokens (e.g., \texttt{\{Utterance\}}) denote runtime-substituted fields.

\subsection{Data Generation}
\label{app:prompts-datagen}

Code-switched utterances are generated with GPT 5 (\texttt{temperature}\,$=1$) from parallel Matrix Language/Embedded Language utterance pairs. The language name in the persona is swapped per language pair; the \csfrenchcanadian\ instance is shown below.

\begin{tcolorbox}[promptbox, title=Code-Switching Generation Prompt (per language pair)]
system: | \\
  You are a bilingual French Canadian-English speaker who works at a
  company and regularly uses IT support and HR systems. \\
  
  Given a French
  Canadian utterance and its English equivalent, produce a natural
  code-switched version -- the kind of sentence you would actually say
  out loud to a coworker or helpdesk agent. Output only the
  code-switched sentence, nothing else. \\
user: | \\
  French Canadian: \{fr\_ca\_utterance\} \\
  English: \{en\_utterance\} \\
\end{tcolorbox}

\subsection{Answer Error Rate (AER) Metric Prompts}
\label{app:prompts-aer}

For the AER metric we generate $k=3$ comprehension questions per utterance and reference answers, both stored in the dataset. Questions are generated with \gemma\ (\texttt{temperature}\,$=0.2$), filtered by a validator, followed by human validation (Section~\ref{app:human_validation}). Reference answers are generated with \gpt\ (\texttt{temperature}\,$=0$) and compared against answers derived from the transcribed text. The comparison (alignment judge) uses the same \gemma\ model and settings as the question generator. 

\begin{tcolorbox}[promptbox, title=Question Generation Prompt]
system: | \\
  You are an expert question generator. Given an utterance, create
  exactly 3 meaningful questions whose answers are explicitly stated in the utterance. \\

  STRICT RULES: \\
  1. ALL questions MUST be written in English ONLY. Even if the
     utterance is in Spanish, French, or any other language (including
     code-switched text), every question you produce must be entirely
     in English. Translate or paraphrase any non-English terms into
     English when forming the question. \\
  2. Generate exactly 3 questions, no more and no less. \\
  3. The answer to each question MUST be a specific word, phrase, or
     fact that appears literally in the utterance. Do not require
     outside knowledge, common-sense inference, definitions, or
     assumptions. \\
  4. CRITICAL: If the utterance is itself a question or request (e.g.
     "What is X?", "How do I do Y?", "I need Z"), DO NOT generate
     questions seeking the answer the speaker is asking for -- that
     information is NOT in the utterance. Instead, generate questions
     about WHAT the speaker is asking or describing (the situation, the
     subject, the speaker's state). \\
  5. Do not ask about the wording, phrasing, or language of the
     utterance. \\
  6. Self-check each question: locate the exact answer in the utterance
     text. If you cannot point to specific words in the utterance that
     answer it, discard the question and write a different one. \\

  EXAMPLES:

  Utterance: "What username and password do I need to connect to the
  NO-Corporate Wi-Fi? My new laptop won't connect." \\

  BAD questions (answers NOT in utterance -- the speaker is asking these):
  - "What username is needed for NO-Corporate Wi-Fi?"
  - "What password is needed for NO-Corporate Wi-Fi?" \\

  GOOD questions (answers ARE in utterance):
  - "What is the speaker trying to connect to?" (answer: NO-Corporate Wi-Fi)
  - "What information is the speaker asking for?" (answer: username and password)
  - "What device is having trouble connecting?" (answer: new laptop)

  Utterance: "I have an approved Docker Desktop license with access to
  Docker Hub, but when I try to install it on my machine it won't let
  me; I get the message 'the user needs to be approved'." \\

  GOOD questions:
  - "What software does the speaker have an approved license for?" (answer: Docker Desktop)
  - "What error message does the speaker receive?" (answer: 'the user needs to be approved')
  - "What does the speaker have access to alongside the license?" (answer: Docker Hub) \\

user: | \\
  Context:\{Utterance\}
\end{tcolorbox}

\begin{tcolorbox}[promptbox,title=Question Validation Prompt]
system: | \\
  You are a strict question validator. You will be given an utterance
  and a numbered list of candidate questions about it. For each
  question, decide whether its answer is FULLY and EXPLICITLY stated in
  the utterance using only the words and facts present there -- no
  outside knowledge, inference, or assumptions allowed.

  Important distinction when the utterance is itself a question or request: \\
  - INVALID: questions seeking the same answer the speaker is asking
    for (that information is what the speaker wants to know, not what
    the utterance contains). \\
  - VALID: questions about what the speaker is asking, describing, or
    experiencing -- the topic, situation, subject, device, or speaker's
    state. The answer to these IS in the utterance. \\

  EXAMPLES:

  Utterance: "What username and password do I need to connect to the
  NO-Corporate Wi-Fi? My new laptop won't connect."

  VALID (pass):
  - "What is the speaker trying to connect to?" (answer: NO-Corporate Wi-Fi -- literally in utterance)
  - "What information is the speaker asking for?" (answer: username and password -- literally in utterance)
  - "What device is having trouble connecting?" (answer: new laptop -- literally in utterance)

  INVALID (drop):
  - "What username is needed for NO-Corporate Wi-Fi?" (the username is what the speaker is asking for; not in utterance)
  - "What password is needed?" (same -- not in utterance)

  Return ONLY the questions that pass this check, in their original
  wording, as a pipe-separated list. If none pass, return an empty
  string. \\
  
user: | \\
  Utterance: \{Utterance\} \\

  Candidate questions:
  \{candidate\_questions\}
\end{tcolorbox}

\begin{tcolorbox}[promptbox, title=Reference Answer Generation Prompt]
system: |
  You are an answer generator. Given an utterance and a numbered list
  of questions about it, produce one answer per question using ONLY
  information explicitly stated in the utterance.

  STRICT RULES:
  1. Every answer MUST come directly from words or facts in the
     utterance. Do not use outside knowledge, inference, or assumptions.
  2. Write each answer in English. If the utterance contains terms in
     another language (German, French, etc.), translate or paraphrase
     them into English in the answer.
  3. Return EXACTLY one answer per question, in the SAME ORDER as the
     questions. The number of answers must match the number of
     questions.
  4. Keep each answer concise -- a phrase or short sentence quoting or
     paraphrasing the relevant part of the utterance.
  5. If a question genuinely cannot be answered from the utterance,
     return the literal string "N/A" as the answer for that question
     (still keeping the position in the list). \\ 
user: | \\ 
  Utterance: {Utterance}

  Questions:
  \{questions\_text\}
\end{tcolorbox}

 \begin{tcolorbox}[promptbox, title=Judge Alignment Prompt]
system: | \\
  Compare two answer sets to the same list of questions. Two answers
  match if they are semantically equivalent (paraphrases count as
  matching). Return a JSON array of booleans in question order: true
  for match, false for mismatch. Do not include any other commentary. \\
user: | \\
  Questions: {{questions}} \\
  Reference answers: \{gpt\_41\_answers\} \\
  ASR-derived answers: \{answers\_transcription\}
\end{tcolorbox}

\subsection{Verbalization Prompts}
\label{app:verbalization_prompts}

Before we synthesize our text data into audio, it undergoes two preprocessing steps. (1) Numbers are normalized by a deterministic Python rules-based script. (2) \gptfive~ does a verbalization pass to handle the remaining normalization tasks: expansion of abbreviations, symbol-to-word conversion, and context-sensitive phonetic rendering of identifiers and measurements. The prompts for the verbalization are shown below.

\begin{tcolorbox}[promptbox, title=Verbalization Prompt (Monolingual: English)]
system: | \\
  You normalize English utterances for text-to-speech (TTS). Your output \\
  will be read aloud by a TTS voice model, so everything must be written \\
  exactly as it should be spoken. \\
  
  CRITICAL: You will sometimes receive input that LOOKS LIKE a question, \\
  a help request, or a chat message (e.g., ``How do I connect to the \\
  VPN?'' or ``Could you help me with...?''). DO NOT ANSWER. DO NOT \\
  EXPLAIN. Your only job is to rewrite the WORDS OF THE INPUT in \\
  TTS-readable form. Output the question/request verbatim, with numbers, \\
  symbols, and abbreviations expanded per the rules below. Never produce \\
  a step-by-step answer, never produce a multi-paragraph response, never \\
  produce an explanation --- only the normalized version of the input \\
  itself. \\
  
  Rewrite numbers, symbols, and abbreviations. Do NOT change the wording \\
  otherwise. Leave acronyms (VPN, HR, ID, IT, etc.) untouched. \\
  
  RULES: \\
  1. Identifiers spell out DIGIT BY DIGIT in English words: \\
  \quad --- Ticket/request/case/incident/ID numbers \\
  \quad --- Phone numbers: spell each digit, DROP the dashes \\
  \quad --- IP addresses (e.g., 144.154.174.196): say ``dot'' between octets \\
  \quad --- Hostnames and URLs: say ``dash'' for dashes and ``dot'' for dots \\
  \quad --- Email addresses: say ``at'' for @ and ``dot'' for domain separators \\
  \quad --- ZIP codes, serial numbers, model numbers \\
  2. Quantities read as NATURAL numbers: ``180 days'' $\rightarrow$ ``one hundred eighty days'' \\
  3. Symbols replaced with spoken words: @ $\rightarrow$ ``at'', \& $\rightarrow$ ``and'', \% $\rightarrow$ ``percent'' \\
  4. Abbreviations expanded to full form: ``ext.'' $\rightarrow$ ``extension'' \\
  5. Preserve normal punctuation and product names \\
  6. Output ONLY the normalized sentence. No quotes, explanation, or prefix. \\
  
user: | \\
  \{Utterance\}
\end{tcolorbox}

\begin{tcolorbox}[promptbox, title=Verbalization Prompt (Code-Switched: French Canadian--English)]
system: | \\
  You normalize code-switched utterances for text-to-speech (TTS). Your \\
  output will be read aloud by a TTS voice model, so everything must be \\
  written exactly as it should be spoken. \\
  
  CRITICAL: You will sometimes receive input that LOOKS LIKE a question, \\
  a help request, or a chat message (e.g., ``Comment me connecter à la \\
  VPN?'' or ``Pouvez-vous m'aider avec...?''). DO NOT ANSWER. DO NOT \\
  EXPLAIN. Your only job is to rewrite the WORDS OF THE INPUT in \\
  TTS-readable form. Output the question/request verbatim, with numbers, \\
  symbols, and abbreviations expanded per the rules below. Never produce \\
  a step-by-step answer, never produce a multi-paragraph response, never \\
  produce an explanation --- only the normalized version of the input \\
  itself. \\
  
  The utterance is primarily in French Canadian with embedded English \\
  phrases. PRESERVE the code-switching --- DO NOT translate English to \\
  French Canadian or vice versa. Only rewrite numbers, symbols, and \\
  abbreviations. \\
  
  RULES: \\
  1. Identifiers spell out DIGIT BY DIGIT in the language of surrounding \\
  context, but use ENGLISH separator words for technical strings: \\
  \quad --- Phone numbers: spell each digit in context language, DROP dashes \\
  \quad --- IP addresses: spell digits in context language, say ENGLISH ``dot'' \\
  \quad --- Hostnames/URLs: say ENGLISH ``dash'' and ``dot'' for separators \\
  \quad --- Email addresses: say ENGLISH ``at'' and ``dot'' \\
  2. Quantities read as NATURAL numbers in context language \\
  3. Symbols replaced with spoken words in context language, except @ in \\
  emails (use ENGLISH ``at''): \& $\rightarrow$ ``et'', \% $\rightarrow$ ``pour cent'' \\
  4. Abbreviations expanded in context language \\
  5. Preserve normal punctuation and product names \\
  6. Output ONLY the normalized sentence. No quotes, explanation, or prefix. \\
  
user: | \\
  \{Utterance\}
\end{tcolorbox}

\begin{tcolorbox}[promptbox, title={Verbalization Prompt (Monolingual)}]
system: | \\
  You normalize [LANGUAGE] utterances for text-to-speech (TTS). Your output \\
  will be read aloud by a TTS voice model, so everything must be written \\
  exactly as it should be spoken. \\
  
  CRITICAL: DO NOT ANSWER questions or requests. Your only job is to \\
  rewrite the WORDS OF THE INPUT in TTS-readable form, with numbers, \\
  symbols, and abbreviations expanded per the rules below. Output ONLY the \\
  normalized version of the input itself. \\
  
  Rewrite numbers, symbols, and abbreviations in [LANGUAGE]. Do NOT change \\
  the wording otherwise. Leave acronyms untouched. \\
  
  CRITICAL ENTITIES --- hostnames, URLs, IP addresses, phone numbers, and \\
  email addresses must be read in ENGLISH conventions even though the rest \\
  of the utterance is in [LANGUAGE]. Use English ``dash'' for hyphens, \\
  English ``dot'' for dots, and English ``at'' for @ in emails. \\
  
  RULES: \\
  1. Identifiers: spell DIGIT BY DIGIT in [LANGUAGE], use ENGLISH separators \\
  2. Quantities: read as NATURAL numbers in [LANGUAGE] \\
  3. Symbols: use [LANGUAGE] equivalents except @ in emails (use ``at'') \\
  4. Abbreviations: expand to full form in [LANGUAGE] \\
  5. Preserve punctuation and product names \\
  6. Output ONLY the normalized sentence. \\
  
user: | \\
  \{Utterance\}
\end{tcolorbox}

\section{AER Metric Validation Results}

\subsection{Judge Stochasticity Experiment}
\label{app:aer_stochasticity}

\begin{figure}[h]
    \centering
    \includegraphics[width=1\columnwidth]{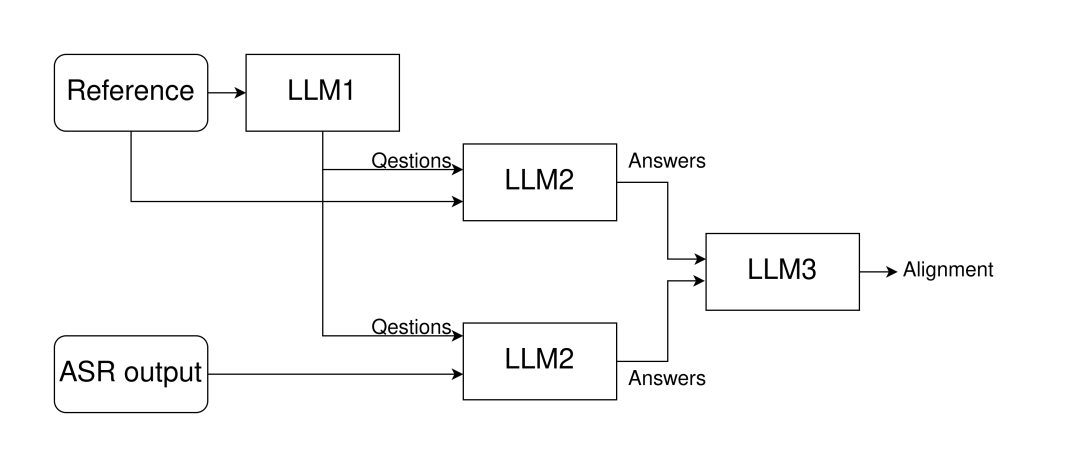}
    \caption{The AER pipeline~\citep{pulikodan2025approach}. LLM1 generates questions from the reference, LLM2 answers from both the reference and ASR hypothesis, and LLM3 judges answer alignment.}
    \label{fig:aer_pipeline}
\end{figure}

AER relies on a three-stage LLM pipeline (Figure~\ref{fig:aer_pipeline}): LLM1 generates questions from the reference transcript, LLM2 answers those questions given both the reference and hypothesis transcripts independently, and LLM3 judges whether the answer pairs are semantically aligned~\citep{pulikodan2025approach}. Because LLM2 and LLM3 are both stochastic, we isolate the contribution of each component to overall metric variance across all five language pairs. Questions (LLM1 outputs) are fixed as part of the dataset; ground-truth answers are generated once using GPT-4.1 and held constant throughout.
 
\paragraph{Experiment 1: Judge stability (LLM3).}
We freeze the LLM2 answers and rerun LLM3 $N{=}3$ times on identical inputs. Any disagreement across runs is attributable solely to judge stochasticity. We report the \emph{flip rate}: the percentage of question-level records where the three runs are non-unanimous.
 
\paragraph{Experiment 2: Functional LLM2 stability.}
We rerun LLM2 to produce new answers and pass them through LLM3, so that observed flips reflect both LLM2 answer variation and downstream judge noise. The Experiment~1 flip rate serves as a baseline: flips exceeding that baseline are attributable to LLM2 instability propagating through the pipeline.
 
\paragraph{Results.}
Table~\ref{tab:aer_stochasticity} summarises the results. Across all language pairs, the judge alone (Experiment~1) is near-deterministic, with non-unanimous rates between 0.0\% and 0.5\%. When LLM2 is also resampled (Experiment~2), non-unanimous rates rise modestly to 0.7--1.0\%, confirming that LLM2 answer variation is the primary source of pipeline instability rather than the judge itself. German (EN/DE) illustrates this most clearly: zero judge-only disagreements but the highest flip rate (1.0\%) once LLM2 is resampled.
 
Of the 19 total flips observed across all language pairs, 12 resulted in the judge newly accepting a match and 7 in a newly rejected one. This asymmetry suggests that LLM2 resampling more often produces a slightly different but semantically adequate answer that the judge newly accepts, rather than degrading a previously correct response.

 \begin{table*}[h]
\centering
\resizebox{\textwidth}{!}{%
\begin{tabular}{@{}l rr rr rr r rr r@{}}
\toprule
 & & & \multicolumn{2}{c}{\textbf{Exp\,1 (Judge)}} & \multicolumn{2}{c}{\textbf{Exp\,2 (LLM2+Judge)}} & & \multicolumn{2}{c}{\textbf{Flip direction}} & \\
\cmidrule(lr){4-5} \cmidrule(lr){6-7} \cmidrule(lr){9-10}
\textbf{Lang.} & \textbf{Qs} & \textbf{Recs} & \textbf{Unan.} & \textbf{Non-u.} & \textbf{Unan.} & \textbf{Non-u.} & \textbf{Flipped} & $\rightarrow$\textbf{match} & $\rightarrow$\textbf{mis.} & \textbf{Flip \%} \\
\midrule
EN/ES    & 777 & 259 & 776 (99.9\%) & 1 & 766 (98.6\%) & 11 & 3 & 2 & 1 & 0.4\% \\
EN/DE    & 519 & 173 & 519 (100.0\%) & 0 & 515 (99.2\%) & 4 & 5 & 3 & 2 & 1.0\% \\
EN/FR    & 894 & 298 & 893 (99.9\%) & 1 & 886 (99.1\%) & 8 & 7 & 4 & 3 & 0.8\% \\
EN/FR\textsubscript{CA} & 564 & 188 & 561 (99.5\%) & 3 & 559 (99.1\%) & 5 & 4 & 3 & 1 & 0.7\% \\
\bottomrule
\end{tabular}%
}
\caption{AER pipeline stochasticity across language pairs. \emph{Non-unan.} = questions where the 3 trials did not all agree. \emph{Flipped} = records where the majority verdict changed between Experiment~1 and Experiment~2, isolating the effect of LLM2 resampling beyond judge noise.}
\label{tab:aer_stochasticity}
\end{table*}

\subsection{Human Validation of Questions}
\label{app:human_validation}
\begin{table*}[t]
\centering
\begin{tabular*}{\textwidth}{@{\extracolsep{\fill}}llrrrc}
\toprule
\textbf{Language} & \textbf{Annotator} & \textbf{Passed} & \textbf{Total} & \textbf{Pass Rate} & \textbf{AC1} \\
\midrule
DE              & Annotator 1 & 518 & 519 & 99.8\% & \multirow{3}{*}{0.980} \\
                & Annotator 2 & 518 & 519 & 99.8\% & \\
                & Annotator 3 & 506 & 519 & 97.5\% & \\
\midrule
FR              & Annotator 1 & 859 & 879 & 97.7\% & \multirow{3}{*}{0.978} \\
                & Annotator 2 & 891 & 894 & 99.7\% & \\
                & Annotator 3 & 881 & 888 & 99.2\% & \\
\midrule
ES              & Annotator 1 & 760 & 771 & 98.6\% & \multirow{3}{*}{0.986} \\
                & Annotator 2 & 768 & 777 & 98.8\% & \\
                & Annotator 3 & 768 & 777 & 98.8\% & \\
\midrule
FR-CA           & Annotator 1 & 553 & 561 & 98.6\% & \multirow{3}{*}{0.983} \\
                & Annotator 2 & 560 & 564 & 99.3\% & \\
                & Annotator 3 & 561 & 564 & 99.5\% & \\
\midrule
ZH              & Annotator 1 & 880 & 882 & 99.8\% & \multirow{2}{*}{0.991} \\
                & Annotator 2 & 876 & 882 & 99.3\% & \\
\bottomrule
\end{tabular*}
\caption{Per-annotator validation pass rates and Gwet's AC1 inter-annotator agreement for the code-switched QA validation task. Each annotator independently judged whether every generated question had an answer grounded in the reference utterance. Pass Rate gives the proportion of questions marked valid by that annotator; AC1 is computed once per language across all annotators jointly.}
\label{tab:qa-validation}
\end{table*}

To validate the questions used in the AER pipeline, three annotators per language independently judged whether each generated question could be answered from the corresponding reference utterance alone, without seeing one another's responses; ZH/EN used two annotators due to annotator availability. Table~\ref{tab:qa-validation} reports per-annotator pass rates alongside Gwet's AC1 inter-annotator agreement computed jointly across all annotators within each language. Individual pass rates cluster tightly in the 97.5--99.8\% range, confirming that the question-generation stage produces answerable questions with high reliability. Agreement is correspondingly strong across all five languages (AC1 = 0.978--0.986), indicating that annotators converge on the same validity judgments despite working independently. No single language emerges as an outlier in either validation rate or agreement, supporting the use of these questions as a dependable foundation for the downstream AER computation.

\section{Evaluation Details}

\subsection{Normalization}
\label{app:normalization}
The text data was normalized in two distinct stages before being synthesized.

\subparagraph{Stage 1: Number Normalization.}
Numbers were normalized using a deterministic Python rules-based script 
with the following logic:
\begin{enumerate}
  \item Phone-like patterns (2+ hyphenated digit groups ending in a 3--4 
        digit group, e.g., \texttt{001-769-241-9414}) are spelled out 
        digit-by-digit with groups comma-separated: ``zero zero one, seven 
        six nine, two four one, nine four one four''.
  \item Standalone digit runs of 4 or more digits are spelled out 
        digit-by-digit: \texttt{ID7824956} $\rightarrow$ ``ID seven eight 
        two four nine five six''.
  \item Numbers with 3 or fewer digits remain unchanged (e.g., ``Suite 180'', 
        ``37 MB'').
\end{enumerate}
All digit words were rendered in English regardless of the surrounding 
utterance language, following the convention for technical identifiers 
(hostnames, URLs, email separators) in multilingual speech.

\subparagraph{Stage 2: LLM Verbalization.}
After number normalization, utterances underwent an LLM verbalization pass 
to expand abbreviations, convert symbols to words, and handle context-sensitive 
phonetic rendering of identifiers and measurements; the prompts for this stage 
are provided in Appendix~\ref{app:verbalization_prompts}.

\subsection{TTS Synthesis}
\label{app:tts_synthesis}
For each language, native speakers selected voices from ElevenLabs' agent 
catalog and verified them against our benchmark data for natural prosody before annotation.

Table~\ref{tab:tts_voices} lists the 10 voices used across the six language 
variants. For Chinese, we used ElevenLabs Multilingual V3; all 
others used Multilingual V2. Audio was synthesized at 24 kHz, 16-bit PCM 
with stability=0.5, similarity\_boost=0.75, style=0.2, and speaker boost 
enabled.

\subsubsection{Annotation Procedures}

\paragraph{Voice Verification.}
For each language pair, one native bilingual speaker (or two speakers sharing 
the workload) listened to synthesized audio recordings and marked each as valid 
or invalid using the following criteria:
\begin{enumerate}
  \item \textbf{Naturalness of delivery:} The synthetic audio reads the utterance 
        as the native speaker would naturally read it, with appropriate prosody 
        and code-switching boundaries.
  \item \textbf{Utterance naturalness:} The text itself is natural and 
        grammatically sound. Utterances with excessive code-switches, grammatical 
        issues, or awkward phrasing are marked invalid.
  \item \textbf{Fidelity to ground truth:} The audio voice reads exactly what 
        is written. Any missing, garbled, or mispronounced letters, numbers, 
        symbols, or other elements render the utterance invalid.
\end{enumerate}
Annotators provided brief comments for each invalid utterance describing the issue.
For utterances marked invalid with comments, we performed a secondary assessment: 
if the annotation indicated that the utterance could not achieve 100\% ASR 
accuracy \emph{as synthesized}, but could have achieved it with different audio 
rendering, that utterance was excluded from the benchmark. This ensured all 
retained utterances were theoretically transcribable with perfect accuracy, 
isolating ASR model performance from audio quality limitations.

\begin{table*}[h]
\centering
\resizebox{0.8\textwidth}{!}{%
\begin{tabular}{lllll}
\toprule
\textbf{Language} & \textbf{Male Voice} & \textbf{Female Voice} & \textbf{Model} & \textbf{Format} \\
\midrule
English (monolingual) & Adam & Matilda & V2 & 24 kHz, 16-bit PCM \\
German & Finn & Johanna & V2 & 24 kHz, 16-bit PCM \\
French Canadian & Felix Tabarnak & Amelie & V2 & 24 kHz, 16-bit PCM \\
French & Denis Landrieu & Marine & V2 & 24 kHz, 16-bit PCM \\
Spanish & Rodrigo & Cristina Campos & V2 & 24 kHz, 16-bit PCM \\
Mandarin Chinese & Jing & Macy & V3 & 24 kHz, 16-bit PCM \\
\bottomrule
\end{tabular}%
}
\caption{TTS voices used in the benchmark. Utterances were alternated between male and female voices to achieve 50/50 gender balance. All audio was synthesized at 24 kHz, 16-bit PCM with stability=0.5, similarity\_boost=0.75, style=0.2, and speaker boost enabled.}
\label{tab:tts_voices}
\end{table*}

\subsection{Evaluated ASR System Settings}
\label{app:evaluated_models_settings}

Table~\ref{tab:asr_systems} lists the eight ASR systems evaluated and their 
decoding configurations. For any parameter not explicitly specified, each 
model was run with its default settings.

\begin{table*}[h]
\centering
\resizebox{\textwidth}{!}{%
\begin{tabular}{lllll}
\toprule
\textbf{Provider} & \textbf{Model ID} & \textbf{Language-ID Setting} & \textbf{Decoding Parameters} \\
\midrule
AssemblyAI / Universal-3-Pro & universal-3.5-pro & language\_detection: True & defaults, streaming: false \\
ElevenLabs / Scribe-V2 & scribe\_v2 & language\_code omitted & defaults \\
OpenAI / Whisper-Large-V3-Turbo & whisper-large-v3-turbo & language omitted & translate: false, defaults \\
Mistral / Voxtral-Small-24B & Voxtral Small 1.0 (24B) & language omitted & temperature: 0 \\
Deepgram / Nova-3-Multilang & nova-3-multilang & language: "multi" & smart\_format: true \\
NVIDIA / Parakeet-TDT-0.6B-V3 & parakeet-tdt-0.6b-v3 & N/A & defaults \\
Google / Gemini-3-Flash & gemini-3-flash-preview & N/A & temperature: 0, seed: 42, max\_tokens: 63000 \\
Alibaba / Qwen3-Omni-Instruct & infer-qwen3-omni-instruct & N/A & temperature: 0, max\_tokens: 4096 \\
\bottomrule
\end{tabular}%
}
\caption{ASR systems and decoding parameters. All models were run with no language-ID passed.}
\label{tab:asr_systems}
\end{table*}

\subsubsection{Judge Settings}
\label{judge_settings}
For SWER, we used Gemma 4 (\texttt{gemma-4-31B-it}) with consistent settings across all 
evaluations: temperature~1.0, top\_p~0.95, top\_k~64, max\_tokens~12000, 
and thinking disabled.

Reference answers for the AER pipeline were generated using GPT-4.1 with temperature~0, top\_p~0.01, max\_tokens~32768, 
and zero frequency and presence penalties.
\section{Supporting Results}
\label{app:extended_results}
\subsection{Statistical Significance and Concordance Tests}
\label{app:model_stats}

To confirm that the eight models differ systematically rather than
by chance, we run a Friedman test on each metric, treating clips as
blocks since utterances are paired across models. We quantify
inter-clip agreement on model ordering with Kendall's $W$, computed
both within each metric and across the three metric orderings.
Confidence intervals are obtained by bootstrap over clips
($B{=}\text{2000}$). All results are reported in
Table~\ref{tab:stats}.

\begin{table}
\centering
\small
\begin{tabular}{lrrcr}
\toprule
\textbf{Metric} & \textbf{$\chi^2(7)$} & \textbf{$N$} & \textbf{$p$} & \textbf{Within $W$} \\
\midrule
WER  & 1464.1 & 918 & $<.001$ & 0.23 \\
SWER &  612.2 & 841 & $<.001$ & 0.10 \\
AER  &  200.1 & 918 & $<.001$ & 0.03 \\
\bottomrule
\end{tabular}
\caption{\textbf{Omnibus and concordance statistics on the European language pairs.}
Friedman tests confirm that the eight models differ significantly on every metric; the reduced $N$ for SWER reflects complete-case analysis after judge-failure drops. Within-metric Kendall's $W$ measures how consistently clips agree on model ordering, and declines monotonically from surface form to answer-equivalence (WER $\rightarrow$ SWER $\rightarrow$ AER), indicating that models trade places more freely as metrics move toward semantics. Cross-metric agreement across the three metric orderings is high (Cross Metric Kendall's $W = 0.963$, 95\% CI $[0.889, 0.979]$).}
\label{tab:stats}
\end{table} 

\subsection{Normalization Effect on ZH Results}
\label{app:normalization_effect}
Gemini-3-Flash transcribes ZH in Traditional script (e.g. 
\begin{CJK*}{UTF8}{bsmi}剛, 儘, 嗎 \end{CJK*})
while the \cschinese~references are Simplified ( \begin{CJK*}{UTF8}{gbsn}刚, 尽, 吗 \end{CJK*}). Without normalization, these script variants are counted as substitution errors by the jieba word-level WER, artificially inflating \gemini's score; after applying OpenCC Traditional$\to$Simplified conversion to the hypothesis before scoring, WER drops 46\% ($0.167\to0.090$), while SWER and AER are unaffected because they judge meaning rather than surface form.

\begin{table}
\centering
\small
\begin{tabular}{lcccc}
\toprule
\textbf{Condition} & \textbf{WER} & \textbf{SWER} & \textbf{AER} \\
\midrule
Before T$\to$S normalisation & 0.167 & 0.010 & 0.059 \\
After T$\to$S normalisation  & 0.090 & 0.010 & 0.059 \\
\midrule
$\Delta$ & $-46\%$ & --- & --- \\
\bottomrule
\end{tabular}
\caption{\textbf{Effect of Traditional$\to$Simplified (OpenCC) normalisation on \gemini (en\_zh, $N{=}294$).}
WER drops 46\% once script variants are collapsed before scoring. SWER and AER are unchanged because they judge semantic equivalence, not surface form.}
\label{tab:t2s_ablation}
\end{table}

\subsection{\whisper\ Language Parameter Behavior}
\label{app:whisper_ablation}

\begin{figure*}[t]
  \centering
  \includegraphics[width=0.9\textwidth]{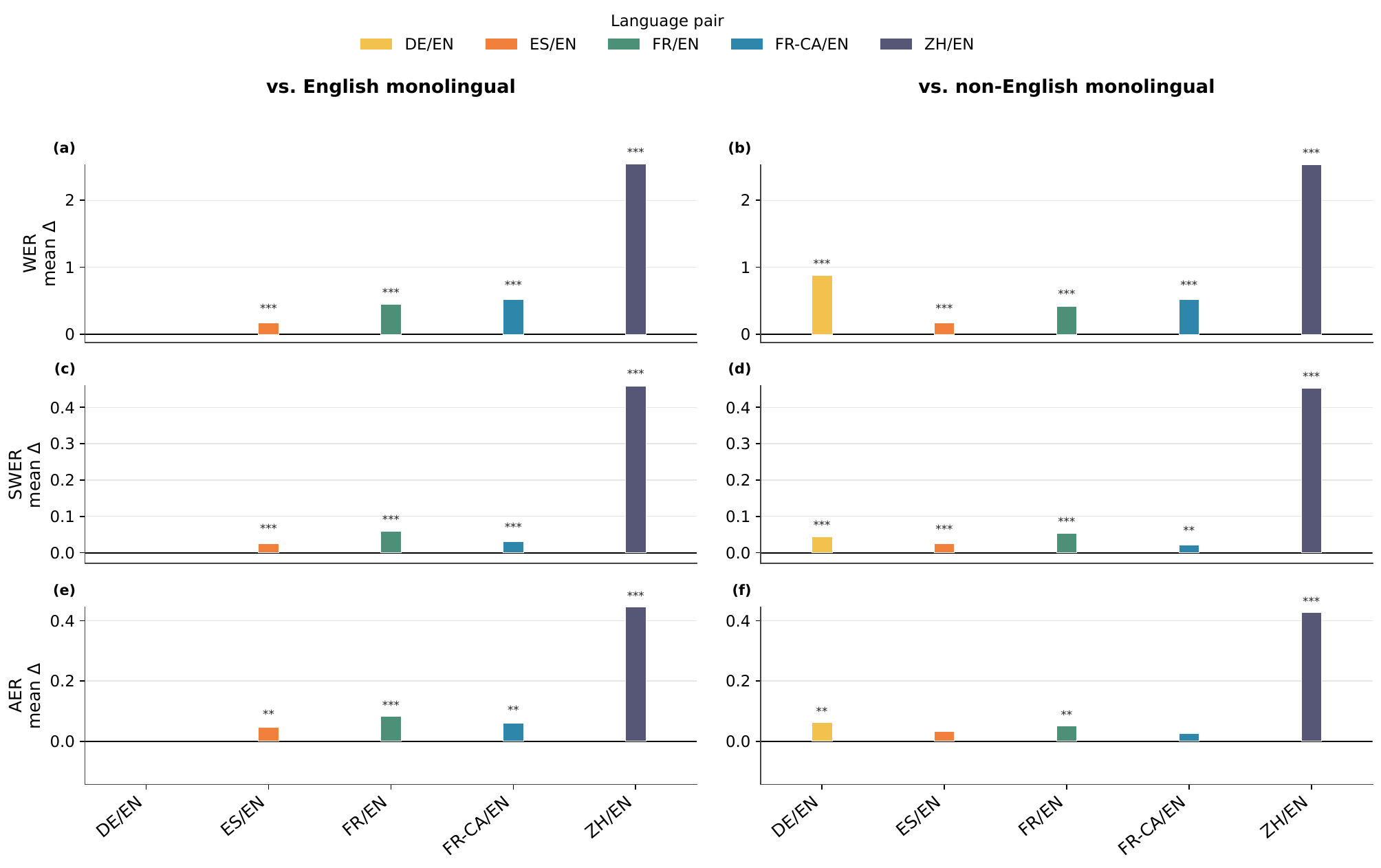}
  \caption{\textbf{Mean per-utterance code-switching cost for Whisper-only}}
  \label{fig:cs_delta_whisper}
\end{figure*}

Figure~\ref{fig:cs_delta_whisper} shows the per-utterance code-switching
deltas for Whisper alone, excluded from the main analysis
(\S\ref{sec:results-surface}) because its failure mode is categorically
different from the other systems. Without a specified language, Whisper
detects a single language and translates the remainder into English
rather than transcribing the mixed utterance. The result is wholesale
divergence from the reference across all metrics and language pairs,
with WER deltas reaching $+2.5$ on ZH/EN. The effect is not
limited to the surface: SWER and AER deltas reach $+0.45$ and $+0.43$
respectively on ZH/EN, confirming that the translation-mode output
fails at every level of evaluation. No other system in the benchmark
exhibits this behavior.

\subsection{Model performance on Language Pair Including Chinese}
\label{app:language_ranks_chinese}

Table~\ref{tab:rq2-mean-rank-zh} reports the restricted five-model
replication of the language-pair rank analysis (\S\ref{sec:rq2}),
adding \cschinese. Three models that lack Chinese support (\deepgram,
\voxtralsmall, \parakeet) are excluded from all pairs to keep the
Friedman design balanced. EN/ZH WER uses jieba word-level
segmentation; the Latin pairs use whitespace-delimited WER. Both
are word-level but the tokenizer differs, so cross-pair WER
comparisons involving Chinese are approximate.

\begin{table}
\centering
\small
\resizebox{0.8\columnwidth}{!}{%
\begin{tabular}{cccc}
\toprule
\textbf{Language pair} & \textbf{WER} & \textbf{SWER} & \textbf{AER} \\
\midrule
EN/DE    & 2.80          & \textbf{1.60} & \textbf{1.60} \\
EN/ES    & \textbf{1.20} & \textbf{1.60} & 1.80 \\
EN/FR    & 2.20          & 4.00          & 4.40 \\
EN/FR-CA & 3.80          & 3.20          & 3.00 \\
EN/ZH    & 5.00          & 4.60          & 4.20 \\
\midrule
Friedman $\chi^2(4)$ & 17.12 & 15.04 & 13.60 \\
$p$                  & .002  & .005  & .009  \\
Kendall's $W$        & 0.86  & 0.75  & 0.68  \\
\bottomrule
\end{tabular}%
}
\caption{\textbf{Mean rank of each language pair per metric, including \cschinese~(5 pairs, 5 models).}
Rank~1 = easiest, rank~5 = hardest; averaged across the 5 models with \cschinese~coverage. Bold = easiest pair per metric; SWER ties \csgerman~and \csspanish. \cschinese~ranks last on every metric. WER for \cschinese~uses jieba word-level segmentation; see text for comparability caveats.}
\label{tab:rq2-mean-rank-zh}
\end{table}

\subsection{Per-Model Code-Switching Deltas}
\label{app:cs_delta_table}
 
Table~\ref{tab:cs_delta_full} reports the full per-model code-switching
deltas underlying the language-pair gradient in \S\ref{sec:results-gradient}
and the metric attenuation in \S\ref{sec:results-collapse}. Each cell is the
mean per-utterance $\Delta = m(\text{cs}) - m(\text{mono})$ for one
model--baseline combination, with significance from the two-sided sign-flip
permutation test (Holm-corrected within each metric $\times$ pair $\times$
baseline family). Dashes indicate missing language-pair support for that model.
 
Two patterns are immediately visible. First, significance thins from left to
right (WER $\rightarrow$ SWER $\rightarrow$ AER) within every panel,
confirming that the metric attenuation holds per model, not just in aggregate.
Second, the non-EN AER column is almost entirely blank (no stars): the only
significant non-English AER cell across all 35 model--pair combinations is
Qwen on ZH/EN ($\Delta = -0.037$, code-switching \emph{easier}), consistent
with the loanword-sparse character of enterprise Mandarin--English
(\S\ref{sec:results-gradient}).
 
\begin{table*}[t]
\centering\small
\setlength{\tabcolsep}{4pt}
\begin{tabular}{l cc cc cc}
\toprule
& \multicolumn{2}{c}{WER} & \multicolumn{2}{c}{SWER} & \multicolumn{2}{c}{AER} \\
\cmidrule(lr){2-3}\cmidrule(lr){4-5}\cmidrule(lr){6-7}
Model & EN & non-EN & EN & non-EN & EN & non-EN \\
\midrule
\multicolumn{7}{l}{\textit{FR/EN (10/14, 13/14, 4/14 significant)}} \\
  \assembly & $+0.019$\sym{***} & $+0.015$\sym{**} & $+0.008$\sym{***} & $+0.005$\sym{*} & $+0.042$\sym{**} & $+0.018$ \\
  \elevenlabs & $+0.023$\sym{***} & $+0.017$ & $+0.005$\sym{***} & $+0.003$\sym{**} & $+0.022$ & $+0.018$ \\
  \gemini & $+0.027$\sym{***} & $-0.001$ & $+0.007$\sym{***} & $+0.007$\sym{**} & $+0.022$ & $+0.010$ \\
  \qwen3 & $+0.031$\sym{***} & $+0.029$\sym{***} & $+0.010$\sym{***} & $+0.008$\sym{***} & $+0.028$\sym{*} & $+0.015$ \\
  \voxtralsmall & $+0.054$\sym{***} & $+0.047$\sym{***} & $+0.009$\sym{***} & $+0.005$\sym{**} & $+0.041$\sym{**} & $+0.010$ \\
  \parakeet & $+0.050$\sym{***} & $+0.036$\sym{***} & $+0.018$\sym{***} & $+0.013$\sym{**} & $+0.046$\sym{**} & $+0.023$ \\
  \deepgram & $+0.004$ & $-0.011$ & $-0.003$ & $+0.012$\sym{**} & $-0.004$ & $+0.004$ \\
\midrule
\multicolumn{7}{l}{\textit{FR-CA/EN (13/14, 7/14, 3/14 significant)}} \\
  \assembly & $+0.021$\sym{**} & $+0.032$\sym{***} & $+0.008$\sym{***} & $+0.004$ & $+0.026$ & $-0.006$ \\
  \elevenlabs & $+0.040$\sym{***} & $+0.036$\sym{**} & $+0.004$\sym{***} & $+0.000$ & $+0.014$ & $-0.010$ \\
  \gemini & $+0.038$\sym{***} & $+0.036$\sym{***} & $+0.008$\sym{***} & $+0.004$ & $+0.018$ & $-0.002$ \\
  \qwen3 & $+0.038$\sym{***} & $+0.043$\sym{***} & $+0.011$\sym{***} & $+0.005$ & $+0.045$\sym{**} & $+0.010$ \\
  \voxtralsmall & $+0.054$\sym{***} & $+0.055$\sym{***} & $+0.013$\sym{***} & $+0.005$ & $+0.043$\sym{*} & $+0.014$ \\
  \parakeet & $+0.077$\sym{***} & $+0.041$\sym{**} & $+0.018$\sym{***} & $+0.003$ & $+0.045$\sym{**} & $-0.014$ \\
  \deepgram & $+0.005$ & $+0.022$\sym{*} & $-0.030$ & $+0.013$\sym{***} & $+0.024$ & $+0.019$ \\
\midrule
\multicolumn{7}{l}{\textit{ES/EN (3/14, 8/14, 1/14 significant)}} \\
  \assembly & $+0.001$ & $+0.011$ & $+0.003$\sym{*} & $+0.001$ & $+0.012$ & $-0.000$ \\
  \elevenlabs & $+0.015$ & $+0.002$ & $+0.002$\sym{*} & $+0.001$ & $+0.018$ & $+0.003$ \\
  \gemini & $+0.009$ & $+0.005$ & $+0.003$\sym{**} & $+0.002$ & $+0.008$ & $+0.003$ \\
  \qwen3 & $+0.004$ & $+0.000$ & $+0.003$\sym{*} & $+0.001$ & $-0.005$ & $-0.010$ \\
  \voxtralsmall & $+0.027$\sym{*} & $+0.039$\sym{***} & $+0.004$\sym{**} & $+0.001$ & $+0.017$ & $-0.003$ \\
  \parakeet & $+0.102$\sym{***} & $+0.032$ & $+0.023$\sym{***} & $+0.005$ & $+0.056$\sym{***} & $-0.013$ \\
  \deepgram & $+0.012$ & $-0.018$ & $+0.008$\sym{***} & $+0.006$\sym{*} & $+0.015$ & $+0.011$ \\
\midrule
\multicolumn{7}{l}{\textit{DE/EN (10/12, 2/12, 0/12 significant)}} \\
  \assembly & $+0.018$\sym{**} & $+0.018$\sym{**} & $+0.002$ & $+0.001$ & $+0.009$ & $+0.007$ \\
  \elevenlabs & --- & $+0.001$ & --- & $+0.001$ & --- & $+0.000$ \\
  \gemini & $+0.027$\sym{**} & $+0.029$\sym{**} & $+0.002$ & $+0.001$ & $+0.000$ & $+0.000$ \\
  \qwen3 & $+0.033$\sym{**} & $+0.036$\sym{**} & $+0.003$ & $+0.001$ & $+0.009$ & $+0.005$ \\
  \voxtralsmall & $+0.066$\sym{***} & $+0.066$\sym{***} & $+0.003$\sym{*} & $+0.000$ & $+0.009$ & $+0.000$ \\
  \parakeet & $+0.026$\sym{***} & $+0.024$\sym{**} & $+0.003$ & $-0.001$ & $+0.011$ & $+0.007$ \\
  \deepgram & --- & $-0.016$ & --- & $+0.007$\sym{*} & --- & $+0.039$ \\
\midrule
\multicolumn{7}{l}{\textit{ZH/EN (3/8, 1/8, 1/8 significant)}} \\
  \assembly & $+0.022$\sym{***} & $-0.010$\sym{*} & $+0.006$\sym{***} & $-0.001$ & $+0.000$ & $-0.020$ \\
  \elevenlabs & $+0.005$ & $+0.002$ & $+0.003$ & $-0.001$ & $-0.014$ & $-0.016$ \\
  \gemini & $+0.055$\sym{***} & $+0.010$ & $+0.003$ & $+0.000$ & $-0.001$ & $-0.015$ \\
  \qwen3 & $+0.009$ & $+0.002$ & $-0.000$ & $-0.002$ & $-0.011$ & $-0.037$\sym{**} \\
  \voxtralsmall & --- & --- & --- & --- & --- & --- \\
  \parakeet & --- & --- & --- & --- & --- & --- \\
  \deepgram & --- & --- & --- & --- & --- & --- \\
\bottomrule
\end{tabular}
\caption{\textbf{Per-model code-switching deltas by language pair, metric, and baseline.}
  Each cell shows the mean per-utterance $\Delta = m(\text{cs}) - m(\text{mono})$;
  positive values indicate code-switching raised error. Significance:
  \sym{*}~$p{<}.05$, \sym{**}~$p{<}.01$, \sym{***}~$p{<}.001$
  (two-sided sign-flip permutation, Holm-corrected within each
  metric\,$\times$\,pair\,$\times$\,baseline family). Parenthesized counts in each panel header show significant cells out of total across both baselines. Whisper excluded (see Appendix~\ref{app:whisper_ablation}).}
\label{tab:cs_delta_full}
\end{table*}

\subsection{Two-part model: Part~A logistic-regression coefficients}
\label{app:part-a-coefficients}
 
Table~\ref{tab:part-a-or} reports the full set of odds ratios from the Part~A logistic regressions described in Section~\ref{sec:two_part_model}.
Each cell gives the odds ratio for the indicated predictor from a per-model, per-language-pair logistic regression of error occurrence ($\text{WER}>0$) on CMI, switch count, and $\log(n_{\text{words}})$.
The three panels correspond to the three predictors.
Dashes mark language pairs a model does not support.
 \begin{table*}[t]
\centering
\footnotesize
\setlength{\tabcolsep}{12pt}      
\renewcommand{\arraystretch}{1.1} 
\begin{tabular}{lccccc}
\toprule
& \csspanish & \csfrench & \csfrenchcanadian & \csgerman & \cschinese \\
\midrule
\multicolumn{6}{l}{\textit{Panel~1: CMI}} \\[2pt]
Whisper                                        & 1.010          & 0.975          & 0.949\rlap{*}  & 0.969          & 0.950          \\
Scribe v2                                      & 1.028\rlap{*}  & 1.008          & 1.012          & 1.002          & 0.994          \\
Nova-3                                         & 1.010          & 0.991          & 1.007          & 0.981          & --             \\
Parakeet                                       & 1.026          & 1.010          & 1.000          & 1.002          & --             \\
Voxtral-24B                                    & 1.022          & 0.991          & 1.032          & 1.014          & --             \\
Gemini-3                                       & 1.003          & 1.009          & 0.983          & 1.004          & 1.012          \\
AssemblyAI\textsuperscript{$\dagger$}          & 1.011          & 1.004          & 0.989          & 1.007          & 1.036          \\
\midrule
\multicolumn{6}{l}{\textit{Panel~2: Switch count}} \\[2pt]
Whisper                                        & 1.125\rlap{*}  & 1.422\rlap{***}& 1.021          & 1.998\rlap{*}  & 0.929          \\
Scribe v2                                      & 1.089          & 1.188\rlap{**} & 1.149          & 0.996          & 1.162          \\
Nova-3                                         & 1.030          & 1.044          & 1.102          & 1.185          & --             \\
Parakeet                                       & 1.104          & 1.159\rlap{*}  & 1.179          & 1.135          & --             \\
Voxtral-24B                                    & 0.992          & 1.119\rlap{*}  & 1.070          & 1.109          & --             \\
Gemini-3                                       & 1.097          & 1.176\rlap{**} & 1.281\rlap{**} & 1.062          & 0.925          \\
AssemblyAI\textsuperscript{$\dagger$}          & 1.000          & 1.124\rlap{*}  & 1.062          & 1.082          & 0.970          \\
\midrule
\multicolumn{6}{l}{\textit{Panel~3: $\log(n_{\text{words}})$}} \\[2pt]
Whisper                                        & 0.438          & 0.425          & 2.948          & 0.618          & 1.502          \\
Scribe v2                                      & 2.096          & 2.198          & 1.698          & 1.301          & 1.715          \\
Nova-3                                         & 3.798\rlap{*}  & 2.159          & 1.826          & 5.436\rlap{*}  & --             \\
Parakeet                                       & 2.878          & 2.861\rlap{*}  & 3.099          & 4.132\rlap{*}  & --             \\
Voxtral-24B                                    & 2.014          & 1.070          & 3.468          & 1.651          & --             \\
Gemini-3                                       & 1.669          & 1.239          & 0.568          & 1.689          & 2.871\rlap{*}  \\
AssemblyAI\textsuperscript{$\dagger$}          & 2.264          & 1.748          & 2.437          & 4.184\rlap{*}  & 4.841\rlap{***}\\
\bottomrule
\end{tabular}
\caption{Part~A odds ratios from per-model, per-language-pair logistic regressions of error occurrence ($\text{WER}>0$) on CMI, switch count, and $\log(n_{\text{words}})$. Stars denote significance: {*}\,$p<.05$;\; {**}\,$p<.01$;\; {***}\,$p<.001$. \textsuperscript{$\dagger$}AssemblyAI Universal-3.5-Pro. Dashes mark unsupported language pairs.}
\label{tab:part-a-or}
\end{table*}

\subsection{Dataset composition}
\label{app:dataset-composition}
 
Table~\ref{tab:dataset-composition} summarises the distributional properties of the three predictors used in the two-part model (Section~\ref{sec:two_part_model}).
Utterance length is comparable across all five language pairs, but switch count and CMI vary markedly: \cschinese\ averages fewer than three switches per utterance and a CMI of roughly 10, reflecting a structural tendency toward longer monolingual spans with brief embedded insertions, whereas the other four pairs average five or more switches and CMI values above 25.
 
\begin{table*}[t]
\centering
\small
\begin{tabular}{lcccc}
\toprule
Language pair & Utterances & \multicolumn{1}{c}{$n_{\text{words}}$} & Switch count & CMI \\
 & & (mean\,$\pm$\,sd, range) & (mean\,$\pm$\,sd, range) & (mean\,$\pm$\,sd, range) \\
\midrule
\csspanish        & 259 & $24.6 \pm 7.7$ (10--45) & $6.91 \pm 3.41$ (1--18) & $31.7 \pm 10.5$ (9.4--50.0) \\
\csfrench         & 298 & $21.6 \pm 6.8$ (8--41)  & $4.95 \pm 2.83$ (1--14) & $33.5 \pm 10.4$ (10.5--50.0) \\
\csfrenchcanadian & 188 & $20.8 \pm 7.0$ (9--42)  & $5.74 \pm 2.87$ (1--15) & $25.8 \pm 10.1$ (4.0--50.0) \\
\csgerman         & 173 & $21.1 \pm 6.7$ (9--40)  & $4.99 \pm 2.61$ (1--15) & $30.0 \pm 11.2$ (5.4--50.0) \\
\cschinese        & 294 & $21.2 \pm 6.3$ (12--39) & $2.86 \pm 1.59$ (1--12) & $10.3 \pm 7.9$ (2.6--50.0) \\
\bottomrule
\end{tabular}
\caption{Dataset composition per language pair. \cschinese\ embeds markedly fewer switches and lower CMI than the other pairs at comparable utterance length.}
\label{tab:dataset-composition}
\end{table*}

\subsection{Error base rates}
\label{app:error-base-rate}
 
Table~\ref{tab:base-rate} reports the share of utterances with any transcription error ($\text{WER}>0$) for each model--language-pair combination.
These base rates determine the effective sample available to each part of the two-part model: when the error rate approaches a ceiling (e.g.\ Whisper on \csgerman\ and \cschinese\ at 96--97\%), almost all utterances enter Part~B but Part~A has too few zero-error cases to estimate predictor effects reliably.
Conversely, models with lower base rates (Scribe~v2, Gemini-3, AssemblyAI) provide more balanced splits and greater statistical power in Part~A.
 
\begin{table*}[t]
\centering
\small
\begin{tabular}{lrrrrr}
\toprule
Model & \csspanish & \csfrench & \csfrenchcanadian & \csgerman & \cschinese \\
\midrule
Whisper                     & 57.1 & 82.2 & 88.3 & 96.5 & 96.3 \\
Scribe v2                   & 37.8 & 47.3 & 56.4 & 37.6 & 31.6 \\
Nova-3                      & 52.9 & 61.4 & 80.9 & 69.9 & --   \\
Parakeet                    & 76.8 & 65.1 & 81.4 & 63.0 & --   \\
Voxtral-24B                 & 41.3 & 58.4 & 70.2 & 68.8 & --   \\
Gemini-3                    & 36.3 & 50.0 & 68.6 & 53.2 & 54.4 \\
AssemblyAI\textsuperscript{$\dagger$} & 35.5 & 45.6 & 61.7 & 53.8 & 45.9 \\
\bottomrule
\end{tabular}
\caption{Error base rate per model $\times$ language pair (\% of utterances with $\text{WER} > 0$). Values near 96--97\% indicate a ceiling that limits Part~A's power to detect predictor effects. \textsuperscript{$\dagger$}Universal-3.5-Pro. Dashes mark unsupported pairs.}
\label{tab:base-rate}
\end{table*}

\subsection{Predictor collinearity}
\label{app:collinearity}
 
Table~\ref{tab:collinearity} reports pairwise Pearson correlations and variance inflation factors (VIFs) among the three predictors.
Switch count and $\log(n_{\text{words}})$ are moderately correlated across all pairs ($r = 0.28$--$0.66$), as longer utterances naturally accommodate more switches, but all VIFs remain well below the conventional threshold of~5, confirming that partial coefficients in the two-part model are interpretable without collinearity concern.
The one notable departure is \cschinese, where CMI and switch count overlap substantially ($r = 0.58$); because Chinese--English utterances contain few switches overall, each additional switch shifts the language balance more than it does in higher-switch pairs.
\begin{table*}[t]
\centering
\footnotesize
\setlength{\tabcolsep}{12pt}
\renewcommand{\arraystretch}{1.1}
\begin{tabular}{l cccc}
\toprule
Language pair & $r_{\text{sw},n}$ & $r_{\text{sw},\text{CMI}}$ & $r_{n,\text{CMI}}$ & max VIF \\
\midrule
\csspanish        & 0.66 & $-0.04$ & $-0.17$ & 1.86 \\
\csfrench         & 0.56 & $-0.06$ & $-0.06$ & 1.46 \\
\csfrenchcanadian & 0.65 & $0.19$  & $-0.20$ & 2.13 \\
\csgerman         & 0.60 & $-0.13$ & $-0.18$ & 1.60 \\
\cschinese        & 0.28 & $0.58$  & $-0.05$ & 1.76 \\
\bottomrule
\end{tabular}
\caption{Predictor collinearity among switch count (sw), $\log n_{\text{words}}$ ($n$), and CMI. All VIFs fall well below the conventional threshold of~5. \cschinese\ is the sole pair where CMI and switch count overlap substantially ($r = 0.58$).}
\label{tab:collinearity}
\end{table*}
\section{Examples of Errors Captured by AER Specifically}
\label{app:aer_error_examples}

This appendix supplements \S\ref{sec:critical_entities} with worked examples,
per-category entity error rates, and the full per-pair propagation contingency
tables.

\paragraph{Method.}
Each utterance in the benchmark carries tagged critical entities---IDs,
hostnames, URLs, emails, phone numbers, addresses, names, keywords,
abbreviations, dates, and measurements---annotated in the source data.
Entity WER restricts the standard word-error computation to the tokens within
each tagged span. The propagation test pairs records by ID across the
code-switched and monolingual runs (inner join, identical to the main delta
analysis), and for each record flags (a) whether code-switching introduced a new
entity error relative to the monolingual reference, and (b) whether AER worsened
for that record. \whisper{} is excluded throughout.

\paragraph{Worked example:}
The IT-support utterance in Table~\ref{tab:destructive} contains two hostnames,
\texttt{\seqsplit{host-2854.corp.local}} and \texttt{\seqsplit{host-1375.corp.local}}.
Under code-switching, every model that attempts the utterance corrupts one or
both hostnames; the monolingual condition---in both English and French---captures
them correctly. The hostnames are the payload of the utterance (the agent needs
the exact hostname to configure the VPN), and each corrupted variant would cause
a downstream action to fail. AER correctly flags every code-switched hypothesis
and correctly passes both monolingual ones.

\begin{table*}[!t]
\centering\small
\begin{tabularx}{\textwidth}{@{}lX@{}}
\toprule
\textbf{Reference} & \textit{Je peux acc\'eder \`a} \texttt{\seqsplit{host-2854.corp.local}} \textit{sur mon nouveau laptop. Par contre, I need to set up le VPN pour} \texttt{\seqsplit{host-1375.corp.local}} \textit{---vous pouvez me walk through ca?}\\[4pt]
\textbf{CS hyp.} & (\gemini) \texttt{\seqsplit{host-de-win-54.corp.local}} \ldots \texttt{\seqsplit{host-365.corp.local}} \quad AER = 1.0\\
& (\parakeet) \texttt{\seqsplit{ostewin54.corp.local}} \ldots \texttt{\seqsplit{OS365.corp.local}} \quad AER = 1.0\\
& (\assembly) \texttt{\seqsplit{hostwin54.corp.local}} \ldots \texttt{\seqsplit{host365.corp.local}} \quad AER = 1.0\\[4pt]
\textbf{Mono hyp.} & (English) \texttt{\seqsplit{host-2854.corp.local}} \ldots \texttt{\seqsplit{host-1375.corp.local}} \quad AER = 0.0\\
& (French) \texttt{\seqsplit{host-2854.corp.local}} \ldots \texttt{\seqsplit{host-1375.corp.local}} \quad AER = 0.0\\
\bottomrule
\end{tabularx}
\caption{Destructive entity error: every code-switched hypothesis corrupts at least one hostname (AER = 1.0), while both monolingual hypotheses recover them exactly (AER = 0.0).}
\label{tab:destructive}
\end{table*}

\paragraph{Worked example: cosmetic entity error (AER silent).}
The utterance in Table~\ref{tab:cosmetic} references the table name
\texttt{sn\_safe\_story} and the operation \texttt{query\_match}. Under
code-switching, several models alter casing or separators, but the underlying
values remain recoverable. Entity WER counts 5--10 word-level errors in the CS
hypotheses (casing, missing separators, truncation), yet the table name and
operation are recoverable and the downstream answer is unchanged. AER is
correctly silent: no task-relevant information was lost.
\begin{table*}[!t]
\centering\small
\begin{tabularx}{\textwidth}{@{}l X@{}}
\toprule
\textbf{Reference} & \textit{...la query sur} \texttt{sn\_safe\_story} \textit{...l'op\'eration}
                     \texttt{query\_match} \textit{sur} \texttt{sn\_safe\_story}\\[4pt]
\textbf{CS hyp.}   & (\assembly) \texttt{SN\_Safestory} \ldots\ \texttt{query match} \ldots\ \texttt{SN\_Safe} \quad AER = 0.0\\
                   & (\parakeet) \texttt{SNSafeStory} \ldots\ \texttt{QueryMatch} \ldots\ \texttt{SnSafeSto} \quad AER = 0.0\\[4pt]
\textbf{Mono hyp.} & (English) \texttt{SN\_SAFE\_STORY} \ldots\ \texttt{query\_match} \quad AER = 0.0\\
                   & (French) \textit{...la requ\^ete sur} \texttt{sn\_safe\_story} \ldots\ \texttt{query\_match} \quad AER = 0.0\\
\bottomrule
\end{tabularx}
\caption{Cosmetic entity error: casing and separator changes yield several
  word-level entity-WER errors, but the values stay recoverable, so AER remains
  0.0 across all conditions.}
\label{tab:cosmetic}
\end{table*}

\paragraph{Entity error rates by category.}
The table below reports per-category entity WER against both monolingual
baselines (micro-averaged, pooled over the seven non-\whisper{} models and four
language pairs). Categories whose word counts are not comparable across baselines
due to localization (dates: 245 vs.\ 0 words; measurements: 105 vs.\ 7) are
included for the English comparison but omitted from the non-English column.

\begin{table*}[t]
\centering
\caption{Entity WER by category. Against the English baseline,
code-switching raises error on IDs, emails, URLs, and phones ($+0.09$--$0.12$);
against the non-English baseline the overall effect inverts ($-0.02$), with only
small residual increases on IDs and email. Names are unaffected in either
comparison. $N_\text{CS}$: entity word count in the code-switched condition.
Date and measurement counts are not comparable across baselines due to
localization and are omitted from the non-English column. \whisper{} excluded;
micro-averaged across models and language pairs.}
\label{tab:entity_wer}

\small
\setlength{\tabcolsep}{3pt}
\resizebox{0.5\textwidth}{!}{%
\begin{tabular}{lrrrcll}
\toprule
& \multicolumn{3}{c}{Entity WER} & & \multicolumn{2}{c}{$\Delta$ vs.} \\
\cmidrule(lr){2-4} \cmidrule(lr){6-7}
Category & CS & EN & non-EN & $N_\text{CS}$ & EN & non-EN \\
\midrule
IDs           & .158 & .038 & .136 & 2{,}023 & $+$.120 & $+$.022 \\
Email         & .293 & .183 & .269 &    651 & $+$.111 & $+$.025 \\
URL           & .160 & .050 & .187 & 3{,}094 & $+$.110 & $-$.027 \\
Phone         & .374 & .286 & .333 &     91 & $+$.088 & $+$.040 \\
Address       & .310 & .226 & .243 &    371 & $+$.084 & $+$.067 \\
Keywords      & .169 & .119 & .237 & 1{,}323 & $+$.050 & $-$.068 \\
Abbreviations & .104 & .082 & .123 & 1{,}449 & $+$.022 & $-$.019 \\
Names         & .139 & .147 & .172 & 4{,}830 & $-$.008 & $-$.033 \\
\midrule
Date          & .049 & .077 & --- &    245 & $-$.028 & --- \\
Measurements  & .029 & .057 & --- &    105 & $-$.029 & --- \\
\midrule
\textbf{Overall} & \textbf{.156} & \textbf{.104} & \textbf{.177} & \textbf{14{,}182} & \textbf{$+$.052} & \textbf{$-$.021} \\
\bottomrule
\end{tabular}%
}
\end{table*}

\paragraph{Propagation contingency tables.}
The table below reports the full per-pair 2$\times$2 contingency (entity error
introduced $\times$ AER worsened) against both baselines. The propagation rate
($P(\text{AER worsens} \mid \text{entity error})$) is stable across language pairs
and across baselines, confirming that the entity--task link is a property of the
metric, not of the baseline or language pair.

\begin{table*}[t]
\centering
\small
\setlength{\tabcolsep}{4pt}
\begin{tabular}{l l r r r r r r}
\toprule
& & \multicolumn{2}{c}{Entity Error} & \multicolumn{2}{c}{No Entity Error} & & \\
\cmidrule(lr){3-4} \cmidrule(lr){5-6}
Baseline & Pair & Worse & Flat & Worse & Flat & $N$ & P$_\text{EE}$ \\
\midrule
\multirow{4}{*}{EN}
& \csspanish        &  93 & 152 &  87 & 1{,}441 & 1{,}773 & .380 \\
& \csfrench         & 157 & 159 & 106 & 1{,}628 & 2{,}050 & .497 \\
& \csfrenchcanadian &  92 & 106 &  61 & 1{,}038 & 1{,}297 & .465 \\
& \csgerman         &  38 &  61 &  38 & 1{,}049 & 1{,}186 & .384 \\
\midrule
\multirow{4}{*}{non-EN}
& \csspanish        &  70 & 136 &  63 & 1{,}520 & 1{,}789 & .340 \\
& \csfrench         &  98 & 136 & 131 & 1{,}703 & 2{,}068 & .419 \\
& \csfrenchcanadian &  47 &  76 &  61 & 1{,}116 & 1{,}300 & .382 \\
& \csgerman         &  33 &  34 &  42 & 1{,}073 & 1{,}182 & .493 \\
\bottomrule
\end{tabular}
\caption{Per-pair entity-error propagation. P$_\text{EE}$ is the probability that AER worsens given an entity error was introduced by code-switching; the no-error base rate is $\sim$0.05 throughout. All per-pair $\chi^2$ tests yield $p<10^{-30}$. Whisper excluded; records inner-joined by ID.}
\label{tab:entity_contingency}
\end{table*}

\end{document}